%% file: LagooNebula_arxiv.tex
\PassOptionsToPackage{pdfpagelabels=false}{hyperref}

\documentclass[useAMS,usenatbib]{mnras}
\usepackage[utf8]{inputenc}
\usepackage[usenames]{color}
\usepackage{ulem}
\usepackage{graphicx}   
\usepackage[T1]{fontenc}
\usepackage{ae,aecompl}

\usepackage[dvipsnames]{xcolor}

\usepackage{orcidlink} 

\usepackage{newtxtext,newtxmath}

\usepackage[T1]{fontenc}
\usepackage{ae,aecompl}

\usepackage{graphicx}	

\def\apj{ApJ}
\def\apjl{ApJL}
\def\aap{A\& A}

\def\mnras{MNRAS}

\def\apjs{ApJS}

\def\mnras{{MNRAS}}

 \def\massage{{\sc{MassAge}}}
 \def\massagemist{{\sc{MassAge-mist}}}
 \def\massageparsec{{\sc{MassAge-parsec}}}

\input definitions

\title[Two groups and violent relaxation in the LNC] {Unveiling two  expanding stellar groups formed through  violent relaxation in The Lagoon Nebula Cluster. } 
\author[Andrea Bonilla-Barroso, et al. ]{Andrea Bonilla-Barroso\orcidlink{0000-0002-6044-125X}$^{1}$, Javier Ballesteros-Paredes\orcidlink{0000-0002-1081-9445}$^1$, Jesús Hernández\orcidlink{0000-0001-9797-5661}$^2$, 
\newauthor Luis Aguilar\orcidlink{0000-0001-7336-2836}$^2$, and Manuel Zamora-Avilés\orcidlink{0000-0002-2133-9973}$^3$
\thanks{E-mail: a.bonilla@irya.unam.mx}
\\
\\
\\
$^{1}$Instituto de Radioastronom\'ia y Astrof\'isica, UNAM, campus Morelia. PO Box 3-72. 58090. Morelia, Michoac\'an, M\'exico
\\
$^{2}$Universidad Nacional Aut\'onoma de M\'exico, Instituto de Astronom\'ia, AP 106,  Ensenada 22800, BC, México
\\
$^{3}$Instituto Nacional de Astrof{\'i}sica, {\'O}ptica y Electr{\'o}nica, Luis E. Erro 1, 72840 Tonantzintla, Puebla, M{\'e}xico
\\
}

\date{Accepted 2024 February 27. Received 2024 February 27; in original form 2023 January 03}

\pubyear{2022}

\begin{document}
\label{firstpage}
\pagerange{\pageref{firstpage}--\pageref{lastpage}}
\maketitle

\begin{abstract}
The current kinematic state of young stellar clusters can give clues on their actual dynamical state and origin. In this contribution, we use Gaia DR3 data of the Lagoon Nebula Cluster (LNC) to show that the cluster is composed of two expanding groups, likely formed from different molecular cloud clumps. We find no evidence of massive stars having larger velocity dispersion than low-mass stars or being spatially segregated across the LNC, as a whole, or within the Primary group. However, the Secondary group, with 1/5th of the stars, exhibits intriguing features. On the one hand, it shows a bipolar nature, with an aspect ratio of $\sim$3:1. In addition, the massive stars in this group exhibit larger velocity dispersion than the low-mass stars, although they are not concentrated towards the center of the group. This suggests that this group may have undergone dynamical relaxation, first, and some explosive event afterward. However, further observations and numerical work have to be performed to confirm this hypothesis. The results of this work suggest that, although stellar clusters may form by the global and hierarchical collapse of their parent clump, still some dynamical relaxation may take place.

\end{abstract}





\section{Introduction}\label{sec:intro}
The detailed way stars are born is one of the most fundamental unsolved problems in astronomy, and so it is how stellar clusters form. To understand the physical processes that allow the formation of stellar clusters, one has to use numerical simulations and theoretical studies and compare the outcomes with the observational properties that such dynamical and chaotic systems exhibit. \\

Recently, contrasting results were found in different young stellar clusters with similar ages. On the one hand, based on Gaia DR2 data, \citet{Wright_Parker19} found that the massive stars in the Lagoon Nebula Cluster (LNC) exhibit larger velocity dispersion compared to low mass stars, which they interpreted as a consequence of the \citet{Spitzer69} instability. On the other hand, using Gaia EDR3 for the Orion Nebula Cluster (ONC),  \citet{BonillaBarroso+22} found that the velocity dispersion is nearly constant as a function of mass. This difference is relevant because, if real, it is evidence of differences in the formation and/or early dynamical evolution of stellar clusters since there is a possible dynamical mechanism behind each one of those results. On the one hand, stellar collisions\footnote{Through the text, we call stellar collisions to indicate, as it is usual in the stellar dynamics community, not that two stars physically collide, but that their interaction is strong enough to modify their trajectories substantially.} promote spatial segregation in mass and velocity, where massive stars concentrate at the center, with a larger velocity dispersion, while lower mass stars are more dispersed in space and with smaller velocity dispersion \citep[e.g., ][]{Binney_Tremaine08}. The Spitzer instability is an extreme case of this process, as we explain in \S\ref{sec:violentANDcollisional}.
{In contrast,}
if clusters are primarily formed by the collapse of a massive cloud or clump, and then the stellar feedback disperses the parent cloud, temporal fluctuations in the ambient potential scatter stars in energy space. This process, called violent relaxation, results in the same velocity dispersion for all stars {independently of their masses} \citep{Lynden-Bell67}. 

Determining the dynamical state of a stellar cluster is non-trivial. For instance, while collisional relaxation produces mass segregation, it does not discard violent relaxation. Indeed, as \citet{McMillan+07} pointed out, if mass-segregated clusters merge, their mass segregation may survive despite the violent relaxation produced by the merger event. In addition, different numerical results have been found that newborn clusters formed in collapsing molecular clouds tend to form their massive stars at the centermost parts of the cluster  \citep[e.g.,][]{Bonnell+07, Kuznetsova+15}. Fortunately, the velocity dispersion of the stars per mass bin is an observable that better helps us to discriminate between the different relaxation mechanisms occurring in a system. Collisional relaxation produces massive stars with larger velocity dispersion than low-mass stars, and violent relaxation produces similar velocity dispersion for all stars, regardless of their mass.  \\

Although challenging, determining the dynamical and structural properties of young stellar clusters is fundamental to understanding the actual dynamical state of the parent cloud. For example, if clouds were supported by turbulence against gravity, one can expect that stellar clusters would form during many free-fall times. Thus, {\color{black} collisional relaxation may take place, and thus, the massive stars in the cluster will increase their velocity dispersion, a process called dynamical heating}. If, instead, the cluster was formed within one or two free-fall times of the parent cloud, one may expect that the velocity dispersion of its newborn stars is nearly constant due to violent relaxation. \\

Given the apparent contradiction between the results by \citet{Wright_Parker19} and ours \citep{BonillaBarroso+22}, and considering that there are new available Gaia DR3 data, we revisited \citet{Wright_Parker19} data by cross-matching their tables with Gaia DR3. As we will show, the stars in the LNC exhibit a constant velocity dispersion as a function of the mass of the stars. In the process, we have found that the Lagoon Nebula Cluster consists of at least two expanding substructures, the main cluster and another expanding substructure that substantially overlaps with the main one, located in the western part of the whole structure. {\color{black} None of them exhibit signs of collisional relaxation, i.e., they are not mass segregated. However, while massive stars in the Primary group do not exhibit dynamical heating, the velocity dispersion of the massive stars in the Secondary group is clearly larger than that of the low-mass stars, a situation that is unclear how to interpret. We speculate about a possible explosive event in this Secondary group, as the cause for this morphological and kinematical behavior.}

The plan for this work is as follows. In \S\ref{sec:violentANDcollisional}, we briefly present the differences between violent relaxation, collisional relaxation and the Spitzer instability. In \S\ref{sec:method_data}, we describe our methodology. Section \ref{sec:results} presents our results. Finally, in \S\ref{sec:discussion} we present our discussion and conclusions.

 \section{Violent vs. collisional relaxation} \label{sec:violentANDcollisional}


\subsection{Violent relaxation}\label{sec:violent}

During the formation of a stellar cluster, the potential well becomes deeper as the cloud collapses. Since the varying potential performs work on the stars, the energy of individual particles is not conserved \citep{Lynden-Bell67}. Furthermore, the particle can gain or lose energy depending on the orbital phase and the timing of the potential variation. The net effect is that particles are scattered in energy, and as long as the potential keeps varying in time, this random scattering will persist. This process is called violent relaxation, and its timescale is of a couple of dynamical timescales \citep{Binney_Tremaine08}.  \\

One can envisage two possible behaviors for the variation of the gravitational potential during violent relaxation. On the one hand, 
{if the stars produce the potential},
the violent relaxation is self-limiting: the more relaxed the particles become, the fewer fluctuations {are} available for driving relaxation. In contrast, if the potential is dominated by an external agent, such as the collapsing molecular cloud during the formation of the cluster, or the expanding molecular cloud during cloud disruption, violent relaxation will continue as long as the potential keeps fluctuating. \\

A key ingredient is that violent relaxation produces equal velocity dispersion for all particles \citep{Lynden-Bell67}. The reason for this effect is that the mass of the stars is utterly negligible compared to the mass represented by the potential fluctuations. It is important to note that, although this process does not produce energy equipartition between particles (they all have the same velocity dispersion), the whole system does it. Indeed, at the end of a collapsing stage, either gas \citep[e.g., ][]{Vazquez-Semadeni+07} or particle systems \citep[e.g., ][]{Noriega-Mendoza_Aguilar18} will tend globally to virialization, with the total gravitational energy being twice the total kinetic energy of the system \citep[see also][]{Ballesteros-Paredes+20}.

\subsection{Collisional relaxation and the Spitzer instability}\label{sec:collisional}

Another relaxation mechanism relevant in few-body systems is collisional relaxation. When particles undergo strong encounters such that their trajectories are substantially deflected, energy is exchanged between them. Although the specific outcome depends on the details of the encounter, the interchange of energy between particles will tend to statistically produce energy equipartition between them, meaning that more massive particles will end up with less velocity than low-mass particles.  \\

Nonetheless, gravitational systems have negative heat capacity \citep{Antonov85, Lynden-Bell_Wood68}, and thus, the statistical tendency of collisions to produce energy equipartition produces the opposite effect in gravitational systems, i.e., as the system loses energy, it becomes hotter dynamically\footnote{As a consequence, the entropy of a gravitational system is not bound, and thus, an isolated system can never reach equilibrium \citep[see, e.g., ][\S4.10.1]{Binney_Tremaine08}.}.
In order to understand this effect, one can imagine a system of only two types of particles: heavy and light, initially thoroughly mixed and in equilibrium with their own potential. As collisions occur, heavy particles lose velocity. Consequently, lacking support, they fall inward to a tighter new equilibrium configuration with larger velocities. 
{The net effect is that mass and velocity segregation occurs as particles try to get energy equipartition via collisions.}
Heavy particles become more concentrated with larger velocity dispersions than light ones \footnote{A point worth noticing is that spatial segregation should be present for collisions to be responsible for the heavier particles having a larger velocity dispersion. If the latter is observed without the former, collisions are not responsible for the dynamical heating.}. In an extreme version of this effect, the massive stars become so concentrated that they evolve on their own collisional timescale and become dynamically detached from the rest of the cluster. This is the Spitzer instability \citep{Spitzer69}. \\

Numerical simulations have consistently found that stellar clusters produce the dynamical heating of massive stars. Then, it is considered that the Spitzer instability typically does occur \citep[e.g., ][]{Trenti_Marel13, Spera+16, Bianchini+16, Parker_Wright16}, with the last authors arguing that it may occur within a few Myrs for young stellar clusters. \\

Theoretically, the difference between the Spitzer instability and the regular mass segregation is a matter of timescales. Without the Spitzer instability, mass segregation grows on the global collisional timescale. With the Spitzer instability, the heavier stars detach dynamically from the rest of the system and collapse at an accelerated pace, that of their own collisional timescale. Observationally, it will be difficult to disentangle whether a cluster has experienced the \citet{Spitzer69} instability, or just collisional relaxation. Being essentially the same process, they both share the same features, though the latter is more extreme than the former. In what follows, thus, we will just talk about collisional relaxation.  \\

\subsection{Violent vs collisional relaxation}\label{sec:violentVScollisional}

While the characteristic timescale for violent relaxation is of the order of 1-2 free-fall timescales
\begin{equation}
  \tauff = \sqrt{\frac{3\pi}{32 G\rho} }
\end{equation}
with $\rho$ the mean density of the system, and $G$ the gravitational constant, the characteristic timescale for collisional relaxation is the collisional timescale, 
\begin{equation}
  \tau_{\rm coll} = \frac{0.1 N}{\ln{N}} t_{\rm cross}
\end{equation}
%
where $N$ is the number of particles, and $t_{\rm cross}\sim R/\sigma_\upsilon$ the crossing time of the system, given by its size $R$ and velocity dispersion $\sigma_\upsilon$ {\citep{Binney_Tremaine08}} . In general terms, the violent relaxation timescale (1-2 free-fall timescales) is substantially smaller than the collisional relaxation timescale. For instance, the free-fall time of a molecular cloud core at a density of 10\ala4~cm\alamenos3 is of the order of $\tauff\sim$1~Myr. In contrast, the collisional timescale of a young stellar cluster with {$\sim$1000} particles (as the Lagoon or the Orion Nebula clusters) and a crossing time of $\sim 1-3$~Myr is of the order of 14-50~Myr. However, it is worth pointing out that numerical simulations by \citet{Parker+16} of young clusters have shown dynamical heating and ejection of massive stars in $\sim 5-10$~Myr. 

From this perspective, one can expect that young clusters, with ages $\lesssim$5~Myr, should exhibit signs of violent relaxation (constant velocity dispersion) preferentially,
while older clusters may show dynamical heating of their more massive stars.


\section{Methods and data} \label{sec:method_data}


Determining how the velocity dispersion varies with mass bins of the stellar objects is one of the critical goals in determining the origin and dynamical state of stellar clusters. The results reported by \citet{Wright_Parker19} for the LNC are at odds with our results found for the ONC \citep{BonillaBarroso+22}. In principle, different clusters may very well have different physical origins; thus, there is nothing wrong with having massive stars undergoing dynamical heating in the LNC while not in the ONC. However, we notice a few relevant points that may affect the inferred result, specifically in terms of determining the masses of the stars. On the one hand, \citet{Wright_Parker19} assumed a single distance and a single extinction for all the stars in the LNC. Both quantities have substantial variation, as shown in \S\ref{sec:sample2}. On the other hand, 
the more massive bin in the analysis of \citet{Wright_Parker19} contains only a few stars, making it 
{sensitive} to small statistics biases. Finally, the analysis by \citet{Wright_Parker19} was performed with Gaia DR2, and thus, the better precision data of Gaia DR3 may allow us to find features that were hidden in more noisy data. \\



In what follows, thus, we used Gaia DR3 data towards the LNC, as explained below. For now, we just recall that using the parallaxes directly reported by Gaia DR3 can overestimate the distance to the cluster (see \S\ref{sec:sample1}). Consequently, we also apply the zero point bias obtained from the ARI Gaia TAP service\footnote{\tt https://gaia.ari.uni-heidelberg.de/tap} to correct the Gaia DR3 parallaxes. The zero-point bias depends on the position of the star in the sky, its brightness, and its colors \citep{Lindegren+21}. In the following, we use the parallaxes corrected by systematics. \\


We also emphasize that in contrast to \citet{BonillaBarroso+22}, where we measured the velocity dispersion using the standard deviation\footnote{More precisely, in \citet{BonillaBarroso+22} we computed the velocity dispersion as 
\begin{equation}
 \sigma_\upsilon = \sqrt{\sigma_{\upsilon_x}^2 + \sigma_{\upsilon_y}^2}, 
\end{equation}
with $\sigma_{\upsilon_{x,y}}$ the standard deviation of the $x$ ($y-$) velocity, which in turn is computed from the RA (DEC) proper motions and distances of each star, previous zero-point bias correction.}, in the present work, we use the interquartile measurements. The reason is that we want to make {the most straightforward} comparison to the results by \citet{Wright_Parker19}. The results obtained using the standard deviation or the interquartile measurement, though not identical, do not change substantially when one or the other is used, except for bins with small statistics, in which case none of the two is a better estimator anyway. \\

In order to properly compare our results with the previous analysis, in the present work, we define three subsets from the sample, as follows.

\subsection{Sample~1: LNC data cross-matched with Gaia DR3}\label{sec:sample1}

To understand the origin of the apparent large velocity dispersion for the most massive stars in the LNC, we define {\bf Sample~1A} as the cross-match between the original 819 stars 
from \citet{Wright+19} with the Gaia~DR3 catalog. We found 817 stars in the original sample have Gaia~DR3 counterparts within 1\arcsec. However, 23 stars were rejected because they lacked parallaxes or had negative values in Gaia DR3. Thus, {Sample 1A} includes 794 stars.\\

We constrain the sample further by keeping those stars from Sample 1A with good astrometric quality, i.e., those stars with RUWE$<$1.4 \citep{Lindegren+21} and parallax errors better than 20\%. It is important to mention that the RUWE parameter is recommended to evaluate the quality of the astrometric solution. Sources with large RUWE are problematic for the astrometric solution, likely to be non-single stars. This step reduced the sample from 794 to 547. \\

We defined {\bf Sample~1B} by drawing from the 547 stars in the previous list, those stars for which proper motions are within three times their corresponding mean absolute deviations (MAD) from the median value in right ascension (pmra) and declination (pmdec). As a result,
{Sample~1B} includes 502 stars. A cautionary point has to be made here: since we are interested in understanding whether the LNC has undergone some degree of dynamical relaxation, eliminating stars with large proper motions may skew our results towards not finding it. In appendix \ref{sec:AppendixRelax} we show that this is not the case, since the rejected stars are not the most massive, and are not particularly concentrated towards the center of the cluster.

\begin{figure}
\includegraphics[width=\columnwidth]{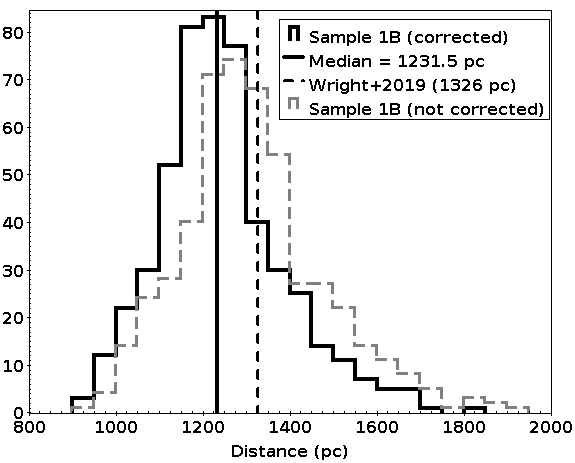}
  \caption{Distances to the stars of Sample~1B. The dashed vertical line denotes the typical value assumed by \citet{Wright+19} and \citet{Wright_Parker19}. The solid line denotes the median distance obtained from the parallaxes corrected by zero-point bias. As a reference, we include the distance distribution of the distance estimated by the parallaxes without the zero-point bias correction.
  }
    \label{fig:distances}
\end{figure}

In Fig.~\ref{fig:distances} we show the distribution of distances of {Sample~1B} based on the parallaxes corrected by the zero point bias (solid line histogram). As a reference, we also include the distance distribution previous to this correction (dashed line histogram). 
Statistically speaking, the distribution of distances to the stars in the LNC is substantially smaller than the typical value adopted of 1.33~kpc in previous studies \citep[e.g., ][]{Kuhn+19, Wright+19, Wright_Parker19}. These differences can affect the derivation of stellar parameters as stellar luminosities and stellar masses. \\

\begin{figure}
\includegraphics[width=0.96\columnwidth]{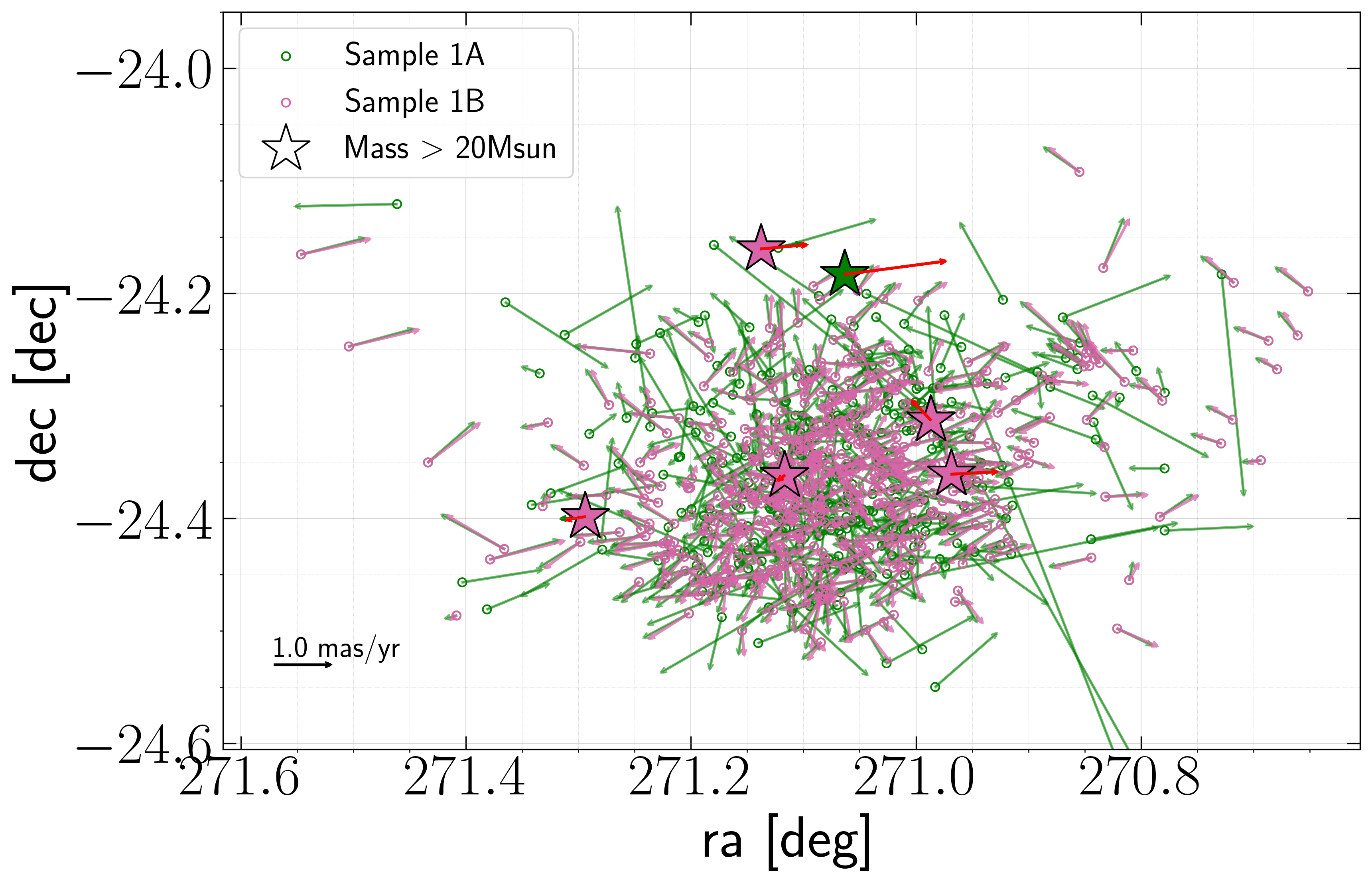} 
  \caption{RA-DEC map of the stars in the LNC, with vectors denoting the proper motions. Green symbols denote the stars in Sample~1A, and pink symbols denote the stars in Sample~1B. Massive stars ($M>$~20~\Msun) are denoted with a star symbol, and their velocity vectors are in red to allow easy distinction from the rest of the velocity vectors. 
  The proper motions of each star in this panel are referenced to the {\color{black} velocity of the centroid} of its corresponding group. 
  }
    \label{fig:mapa:w:propermotions}
\end{figure}

In Fig.~\ref{fig:mapa:w:propermotions}, we show the map of the LNC for Samples~1A (green) and 1B (pink). The proper motions of each sample are denoted with arrows and are plotted in the frame of reference of the whole cluster. For purposes that will be clear later, we also denote with a star symbol those stars with masses larger than 20~\Msun, according to \citet{Wright_Parker19}. 

\subsection{Sample~2: the \massage\ sample}\label{sec:sample2}

We defined {Sample~2} as follows: Since we are interested in an accurate estimation of the masses of the stars,  from Sample~1A we select those stars from \citet{Wright+19} for which effective temperatures have been estimated. We additionally included four stars with effective temperatures estimated by  \citet{Prisinzano+19}, who used similar spectroscopic data and method to \citet{Wright+19}. We also applied the same filters 
as in the case of Sample~1B. While this sample has only 40\%\ of the stars in Sample~1B (286 stars), it still gives us good statistics that may allow to confirm or discard the results found with the masses tabulated in  \citet{Wright_Parker19}.\\

{We estimated extinctions ($A_V$), luminosities, masses, and ages of {Sample 2} using the 
code, \massage{} (Hernández et al. in prep). This code uses as input the effective temperature, the parallaxes, the {Gaia}-DR3 (Gp, Rp, Bp), and 2MASS (J and H) photometry. In brief, $A_V$ is estimated by minimizing the differences between the observed colors corrected by extinction and the theoretical colors obtained from the evolutionary models. We used two evolutionary models in this work, PARSEC \citep{Marigo+17} and MIST \citep{Dotter16}. Luminosities are derived using the extinction-corrected J magnitude, the theoretical bolometric correction, and the distances estimated as the inverse of the parallax.  Finally, stellar masses and ages are obtained by comparing the location of each source on the HR diagram and theoretical evolutionary models. We obtain the mass and age corresponding to specific effective temperature and luminosity values using a linear interpolation method on the theoretical grid.}\\

\begin{figure}
\includegraphics[width=\columnwidth]{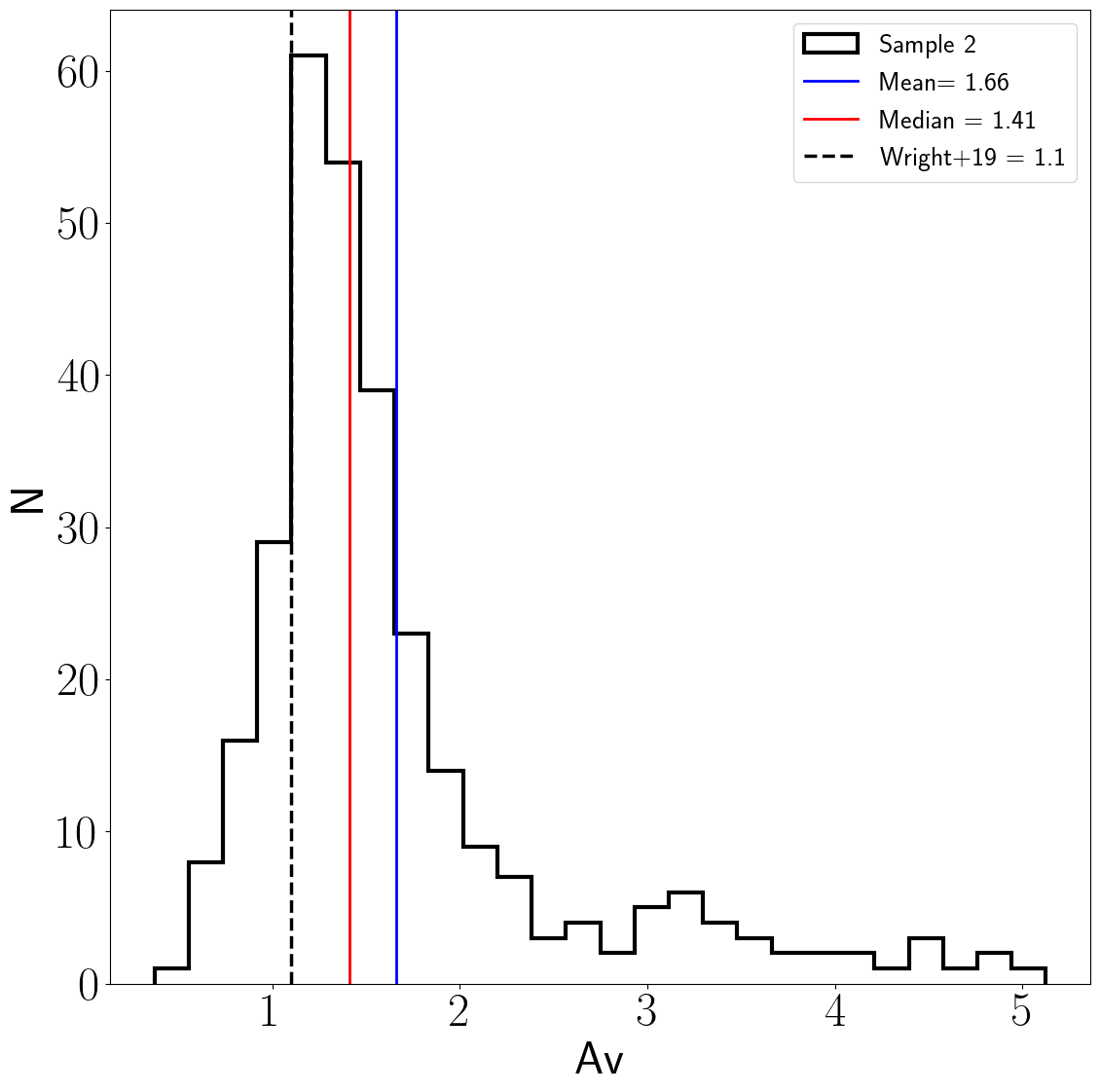}
  \caption{Distribution of extinctions $A_V$ to the stars in Sample~2. The blue vertical line denotes the typical extinction value assumed by \citet{Wright+19} 
 }
    \label{fig:Av}
\end{figure}

In Fig.~\ref{fig:Av} we show the distribution of extinctions towards Sample~2. As can be seen, the extinctions in the sample are statistically larger than the value assumed by \citet[][dashed vertical line]{Wright_Parker19}. {Moreover, a range of extinctions is expected in star-forming regions; thus, assuming a unique value of extinction for a very young stellar population could  produce erroneous results}.\\

\begin{figure}
\includegraphics[width=\columnwidth]{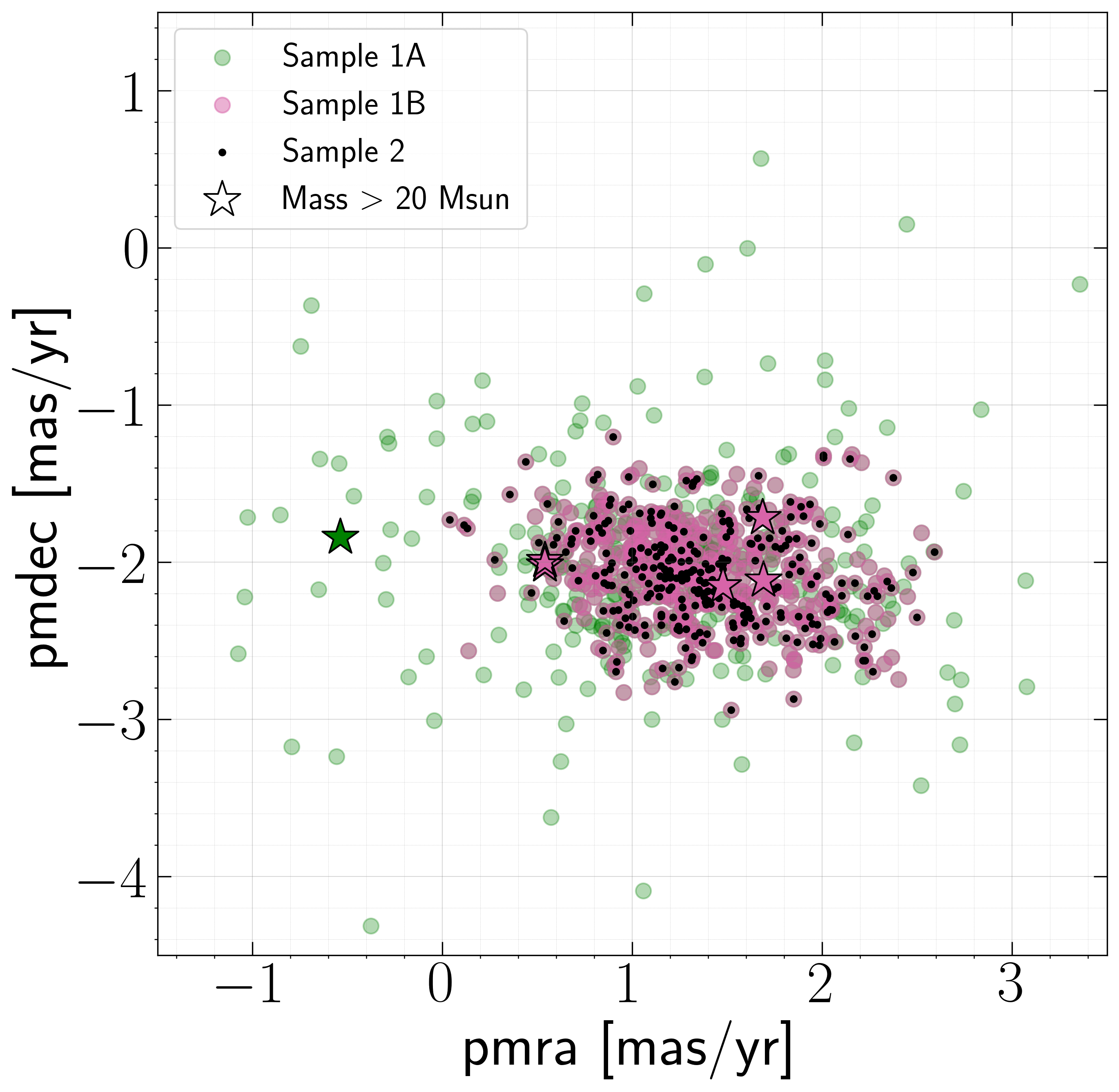} 
  \caption{Vector-point diagram of the stars in the LNC in Sample~1. Green circles denote the proper motions of the stars in Sample~1A, while the pink circles represent those stars in Sample~1B. Black dots denote the proper motions of stars in Sample~2. The six star symbols denote the massive stars ($M>$~20~\Msun). 
}
    \label{fig:punto-vector}
\end{figure}

To show the differences between these samples, in Fig.~\ref{fig:punto-vector}, we show the vector-point diagram, i.e., the proper motion in right ascension {\it vs.} proper motion in declination, for the stars in Sample~1A (green) and 1B (pink). In addition, we show with star symbols the six most massive stars in Sample~1A, i.e., those stars with masses larger than 20~\Msun, according to \citet{Wright_Parker19}. Finally, the black dots denote the stars entering Sample~2, which is basically a subset of Sample~1B, with the addition of those 4 stars from \citet{Prisinzano+19}.  The physics of this plot is discussed in \S\ref{sec:discussion}. Here we just point out that the original sample has many stars which proper motions are substantially more scattered than 3$\sigma$. 

\section{Results}\label{sec:results}
The goal of the present work was to investigate whether the LNC exhibits kinematical signatures of violent or collisional relaxation. In the process, we discovered that the LNC is composed of two expanding groups with a relative velocity of $\sim$3~$\kms$ in the plane of the sky. Thus, we start this section by describing how we did find the two groups, and then we investigate the relaxation nature of the cluster and their groups.

\subsection{Two expanding groups}\label{sec:twogroups}

When looking at kinematical features, one of the first ones are whether a cluster is expanding or not. While it was clear that the LNC is expanding, it was yet to be clear to what extent. Gaia DR2 studies of the LNC have shown contradictory results. On the one hand, \citet{Kuhn+19} reported a moderate expansion.  However, although \citet{Wright+19} confirmed the expansion along the declination, they found no signs of expansion in the right ascension. It is interesting to notice, nonetheless, that visual inspection of Fig.~9 of \citet{Wright+19} hinted at some degree of expansion along the right ascension, which their fit missed, likely because of lower-quality data of Gaia DR2 compared to Gaia DR3. In what follows, we will call ``Primary'' group to the group having most of the stars, and ``Secondary'' to the smaller group.  \\ 

\begin{figure}
  \includegraphics[width=0.95 \columnwidth]{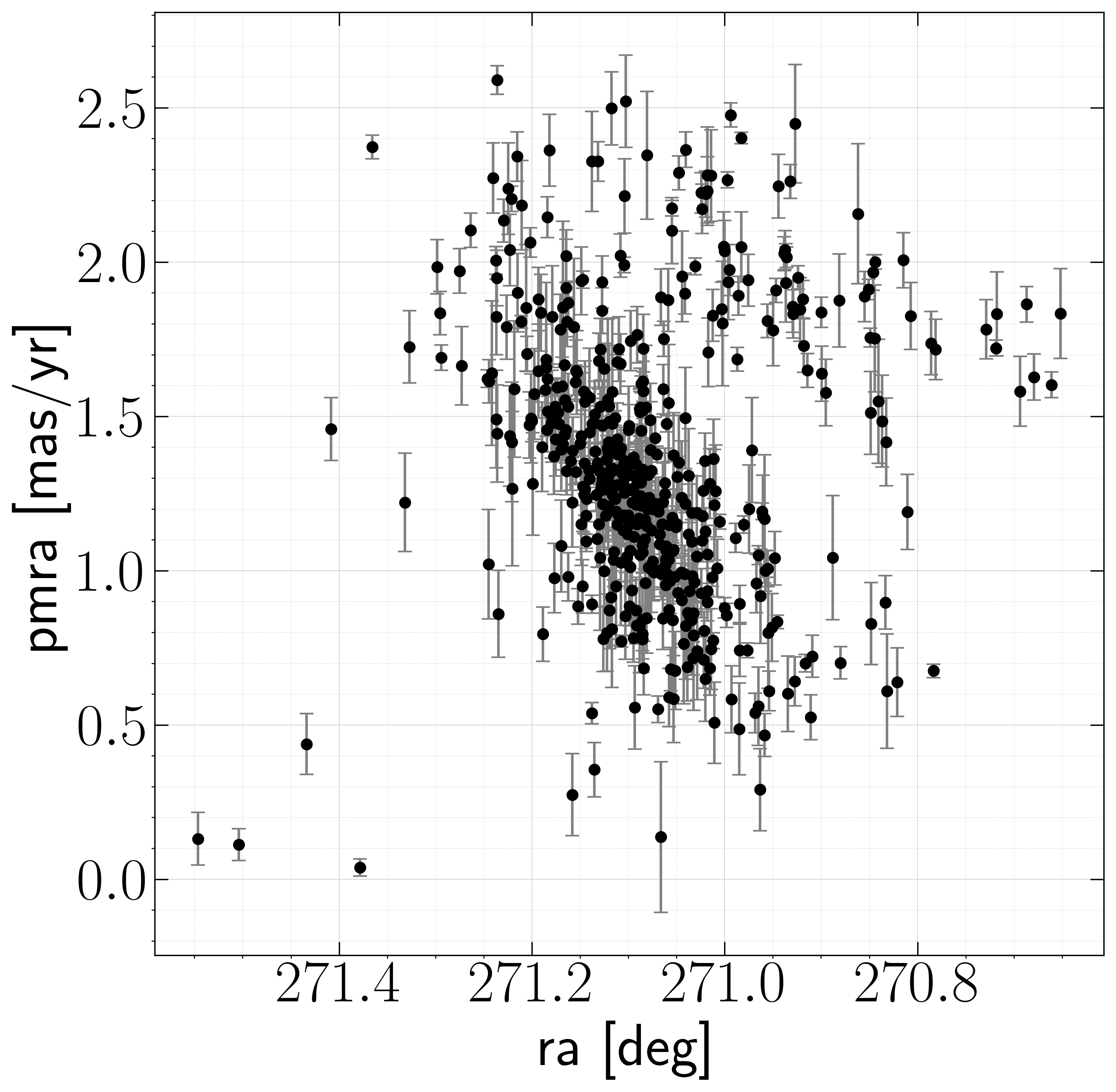}
  \includegraphics[width=\columnwidth]{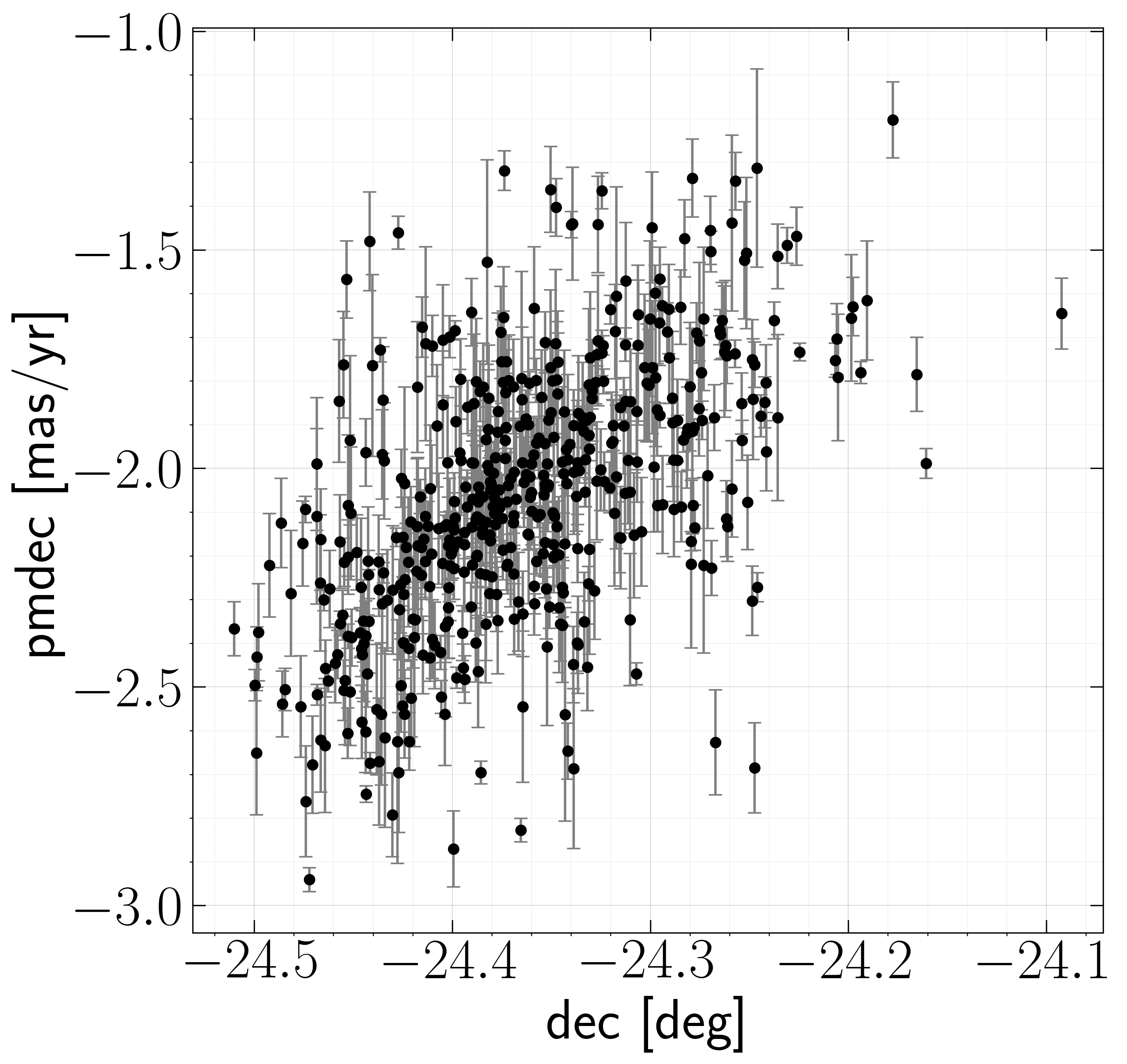} \\
  \caption{Velocity gradients diagrams, (ra, pmra) and (dec, pmdec), for the stars in Sample~1B. The upper panel shows two distinct kinematical groups in the LNC, with characteristic proper motions pmra. These are well mixed, however, in ra, dec and pmdec.}
  \label{fig:pos-vel}
\end{figure}

Aimed to disentangle to which degree the LNC is expanding, in Figure~\ref{fig:pos-vel} we show the (ra,~pmra) and (dec,~pmdec) diagrams of our Sample~1B, which are the equivalent of those in Fig.~9 of \citet{Wright+19}, but for our 3$\sigma$ cleaned Gaia DR3 data  (see \S\ref{sec:method_data}). From this figure, it is clear that the LNC, which has been identified as a well-defined young cluster in the plane of the sky \citep{Walker+56, Adams+83}, is actually composed of two expanding groups that are well-mixed in ra, dec, and pmdec, but clearly separated in the pmra as measured by Gaia DR3. \\

In order to find out which stars belong to each group, it is necessary to apply a clustering method. There are many possibilities to do so, and each one will not necessarily provide the same outcome since the result depends on the method's assumptions. Given the evident velocity gradients in Fig.~\ref{fig:pos-vel}, we decided to apply the K-means algorithm to the (ra,~pmra) plane in order to assign the points to one or the other velocity gradients. The details of our procedure are described in appendix~\ref{sec:appendix}. \\

\begin{figure*}
  \includegraphics[width=0.95 \columnwidth]{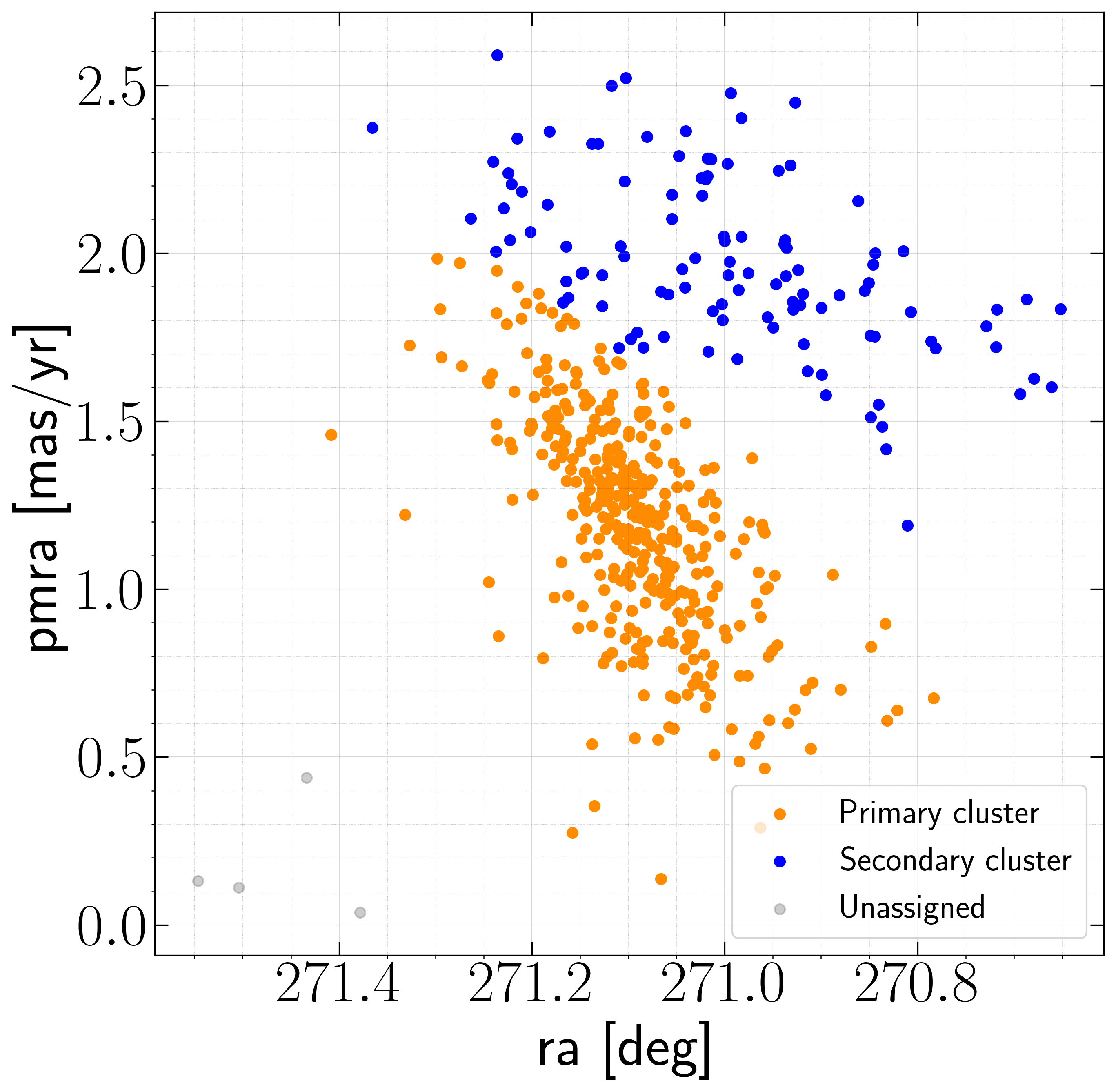}
  \includegraphics[width=\columnwidth]{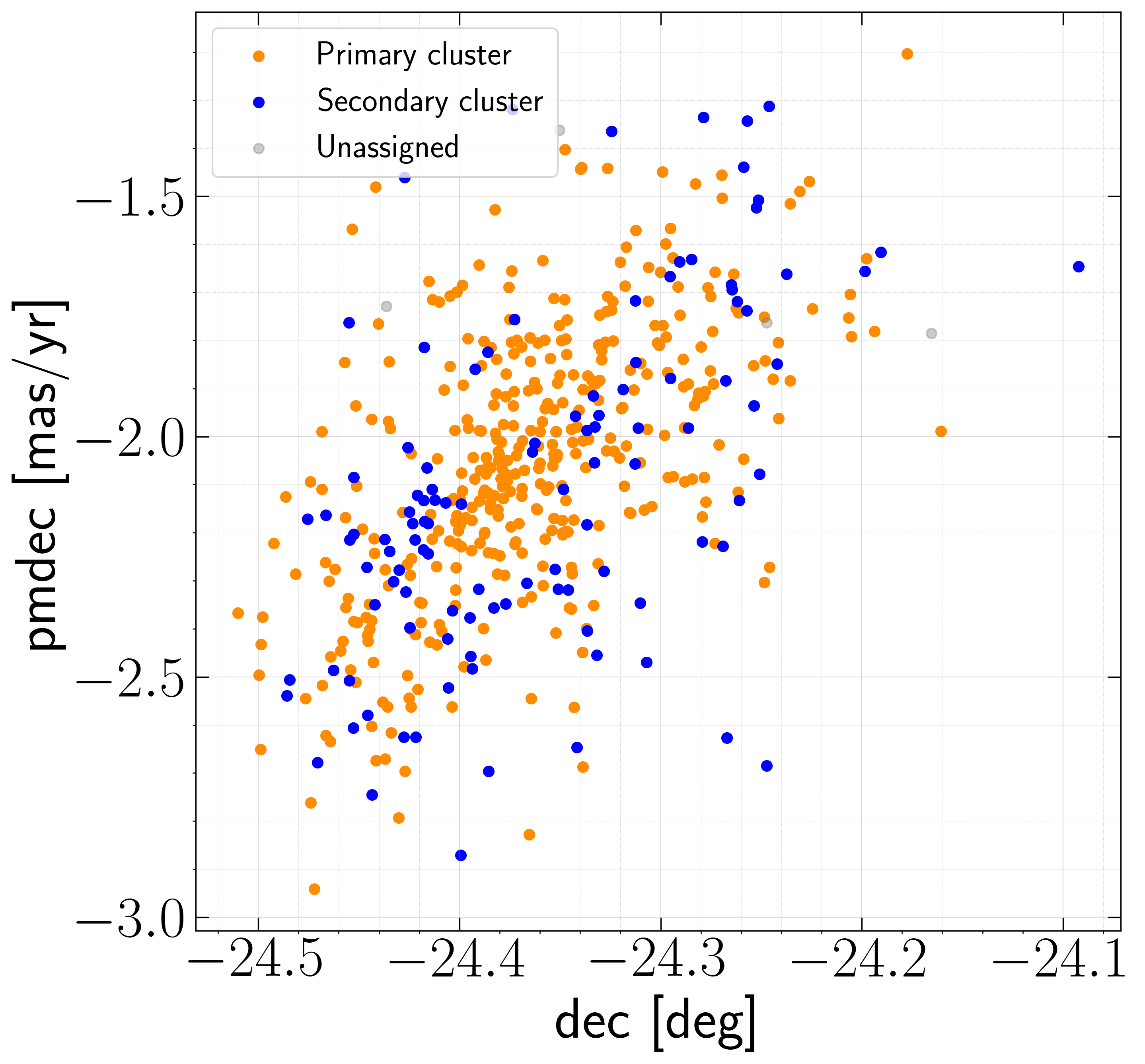}
  \caption{Same as Fig.~\ref{fig:pos-vel}, but color coded according to our cluster-finding algorithm. Orange and blue dots correspond to the Primary and Secondary groups. Grey dots are not assigned to either. Although both clusters have similar expanding rates and are well mixed in declination, they have different rates in right ascension, and the Secondary group is preferentially in the western part of the LNC.}
  \label{fig:pos-vel:new}
\end{figure*}

\begin{figure}
\includegraphics[width=\columnwidth]{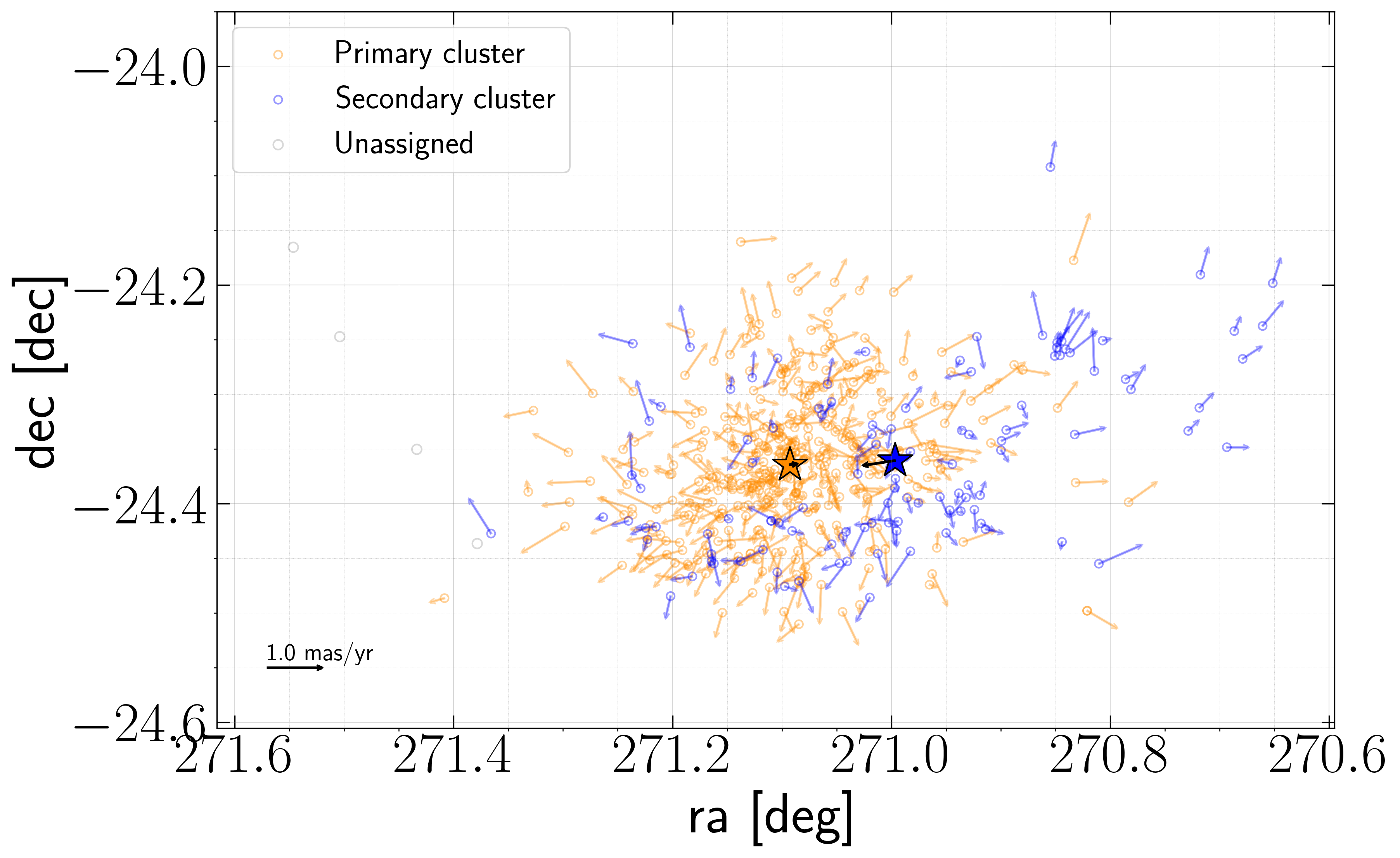}
  \caption{{Ra-dec map of the stars with arrows denoting their proper motions, but denoting stars in the Primary (Secondary) group in orange (blue). The proper motions of each star in this panel are referenced to the center of mass of its corresponding group. It can be noticed that both clusters show expansion motions.  }
   }
    \label{fig:mapa:w:propermotions:new}
\end{figure}

The two groups found with our method are composed of {385 and 110 stars, } respectively. Fig.~\ref{fig:pos-vel:new} shows the (ra,~pmra) and (dec,~pmdec) diagrams of the LNC, with the Primary and Secondary clusters denoted by orange and blue dots, respectively. The (ra,~dec) map with proper motions is shown in Fig.~\ref{fig:mapa:w:propermotions:new}, where the proper motions of each star are shown in the frame of reference of the group they belong to. The big stars denote the position of the centroid of each group, while the corresponding arrows denote their proper motions in the frame of reference of the LNC. {The proper motions difference between the centroids of both groups ($\Delta$pmra$\sim$0.728~mas~yr\alamenos1 and $\Delta$pmdec$\sim$0.106~mas~yr\alamenos1 correspond to a relative velocity between the groups of $\sim$3~km~s\alamenos1 at the median distance of the LNC (d = 1231~pc, see Fig.~\ref{fig:distances}).} \\

In order to compute the expansion rates of each cluster and their dynamical times and to evaluate each group's projected morphology, {we computed the positions and velocities of the stars, assuming they all have the median distance to the LNC.  We associated their $x$ direction to the right ascension and the $y$ direction to the declination. The linear regressions in the space phase ($x$, $v_x$) gave us expanding rates of 0.68~km/s/pc and 0.26~km/s/pc for the Primary and Secondary groups, with Pearson correlation coefficients of $r=$ 0.62 and 0.55 respectively. In the ($y$, $v_y$) space phase, the corresponding expanding rates are 0.69~km/s/pc and 0.68~km/s/pc, with Pearson correlation coefficients of $r=$ 0.57 and 0.55. } From these values, one can notice that the dynamical times for the Primary group are consistent in both directions: {$\tau_{\rm dyn}\sim$1.46 and 1.44~Myr in the ra and dec directions, respectively. Instead, the dynamical timescales for the Secondary group appear less consistent: $\tau_{\rm dyn}\sim$3.89 and 1.47~Myr } for the ra and dec, respectively. All these dynamical ages, however, are consistent with the most ($\sim 90\%$) of the cluster's stars having ages smaller than 10~Myr, with a mean of $2-3$~Myr.\\

This bipolar nature of the Secondary group is consistent with the values in the diagonal of the covariance matrices, which are $var(\bf{x})=10.5$, $var(\bf{y})=2.8$. These values show that the Secondary group is strongly elongated in the $x$ (ra) direction, {\color{black} while the Primary group is substantially less elongated}. Its values of the covariance matrix along the diagonal are $var(\bf{x})=3$, $var(\bf{y})=1.8$.\\

The fact that the LNC comprises two expanding groups has probably passed unadvertised for several reasons. First of all, when plotting the proper motions of the whole cluster, it is not clear that there are two groups, neither in the Sample by \citet[][Sample~1A, green points in Fig.~\ref{fig:mapa:w:propermotions}]{Wright_Parker19} nor in Sample~1B (pink points). Indeed, although the Secondary group is located at the western edge of the main group, it still shares similar positions with the Primary group. In addition, the lack of accurate proper motions even with Gaia~DR2, has made difficult the separation. Finally, while the Primary group is denser at its center, the Secondary group is not strongly centrally concentrated, making it difficult its identification as a group. Instead, it has some  degree of internal structure, since it is composed of several overdensities that have been reported by \citet[][see their Fig.~2b]{Kuhn+14}, with the north-western group being one of the more prominent overdensities.\\

\subsection{Masses of the stars. Comparison between methods}

As commented before, the masses estimated by \citet{Wright+19} are based on assuming a single distance and single extinction. This assumption may give inaccurate mass estimations, potentially changing the main result, namely, that massive stars have larger velocity dispersion. In order to verify that this is not the case, we used our Sample~2, which is based on the crossmatch between Sample~1B and those stars that have reliable estimations of the effective temperature, and thus, that may allow us to estimate more reliably the masses of the stars. \\

\begin{figure*}
\includegraphics[width=0.66\columnwidth]{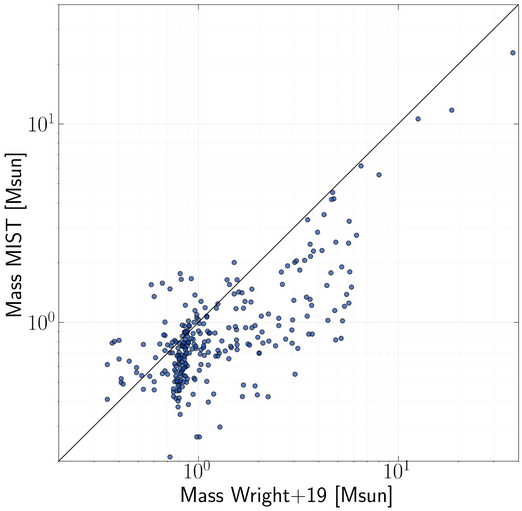}
\includegraphics[width=0.66\columnwidth]{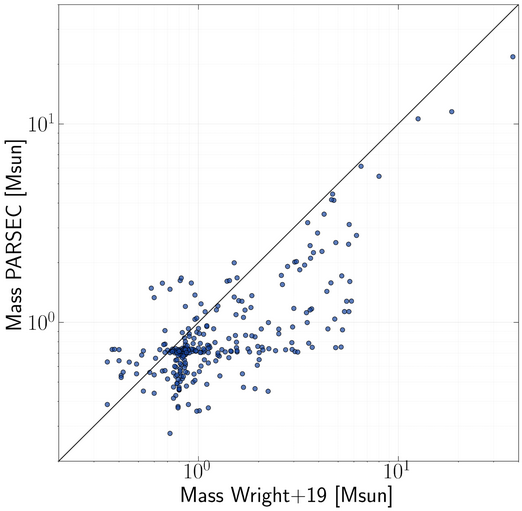}
\includegraphics[width=0.66\columnwidth]{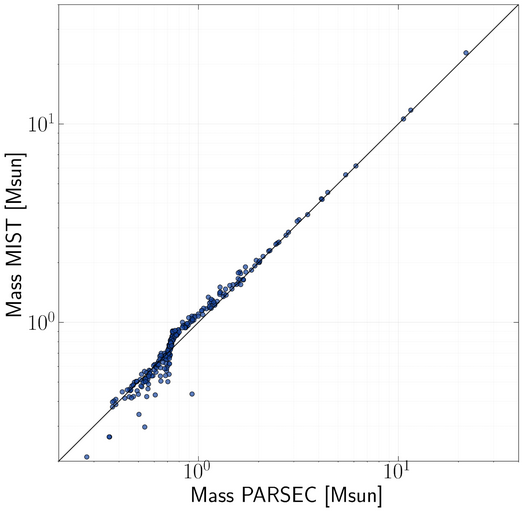}
  \caption{Masses of the stars in the Lagoon Nebula Cluster. (a) masses estimated with \massage\ using the MIST  models of stellar evolution ($y$-axis) {\it vs.} masses reported in \citet{Wright+19} ($x$-axis). (b) masses estimated with \massage using PARSEC models of stellar evolution ($y$-axis) {\it vs.} masses reported by \citet{Wright+19}. (c) masses using \massage: MIST ($y$-axis) {\it vs.} PARSEC ($x$-axis). }
    \label{fig:MassAge}
\end{figure*}

In Fig.~\ref{fig:MassAge} we show the masses estimated with \massage\footnote{The new masses for 286 stars are reported in an electronic Table.}\ and compare them to the masses reported by \citet{Wright+19}. In panels (a)  and (b), we compare the masses estimated by \citet[][$x$ axis]{Wright+19} with the masses estimated with \massagemist{} and \massageparsec, respectively. In panel (c), we compare the differences between our two estimations. The solid lines in each panel are the identity. \\

There are two points to notice from this figure. On the one hand, the masses estimated by \citet{Wright+19} are statistically larger than those estimated with \massage. On the other hand,
differences between the MIST and PARSEC models of early stellar evolution do not appear to give statistically different results. \\

The fact that the masses estimated previously are larger than ours must be because the distance assumed to the stars is statistically larger than the distances from Gaia DR3 corrected by the zero-point bias (see Fig.~\ref{fig:distances} and discussion in \S\ref{sec:method_data}), and thus, the inferred luminosities of the stars will also be larger.

\subsection{No evidence of spatial mass segregation in the Lagoon Nebula Cluster }\label{sec:SpatialSegregation}

It is interesting to investigate whether mass segregation exists in the LNC, especially because although spatial mass segregation does not necessarily imply collisional relaxation, the lack of the former necessarily discards the latter (see \S\ref{sec:collisional}). \\ 

Aimed to quantify the degree of mass segregation in the LNC, we performed KS tests to assess the statistical significance of the difference between the spatial distribution of massive and low-mass stars, using projected distances from the center of mass at the mean distance of the group to determine their spatial position. Since the mass $\mlim$ above which we will distinguish massive from low-mass stars is rather arbitrary, we computed their spatial distributions and applied KS to check differences using different values of $\mlim$ in the range 0.4~\Msun$\leq\mlim\leq$~8~\Msun.\\

\begin{figure*}
\includegraphics[width=0.66\columnwidth]{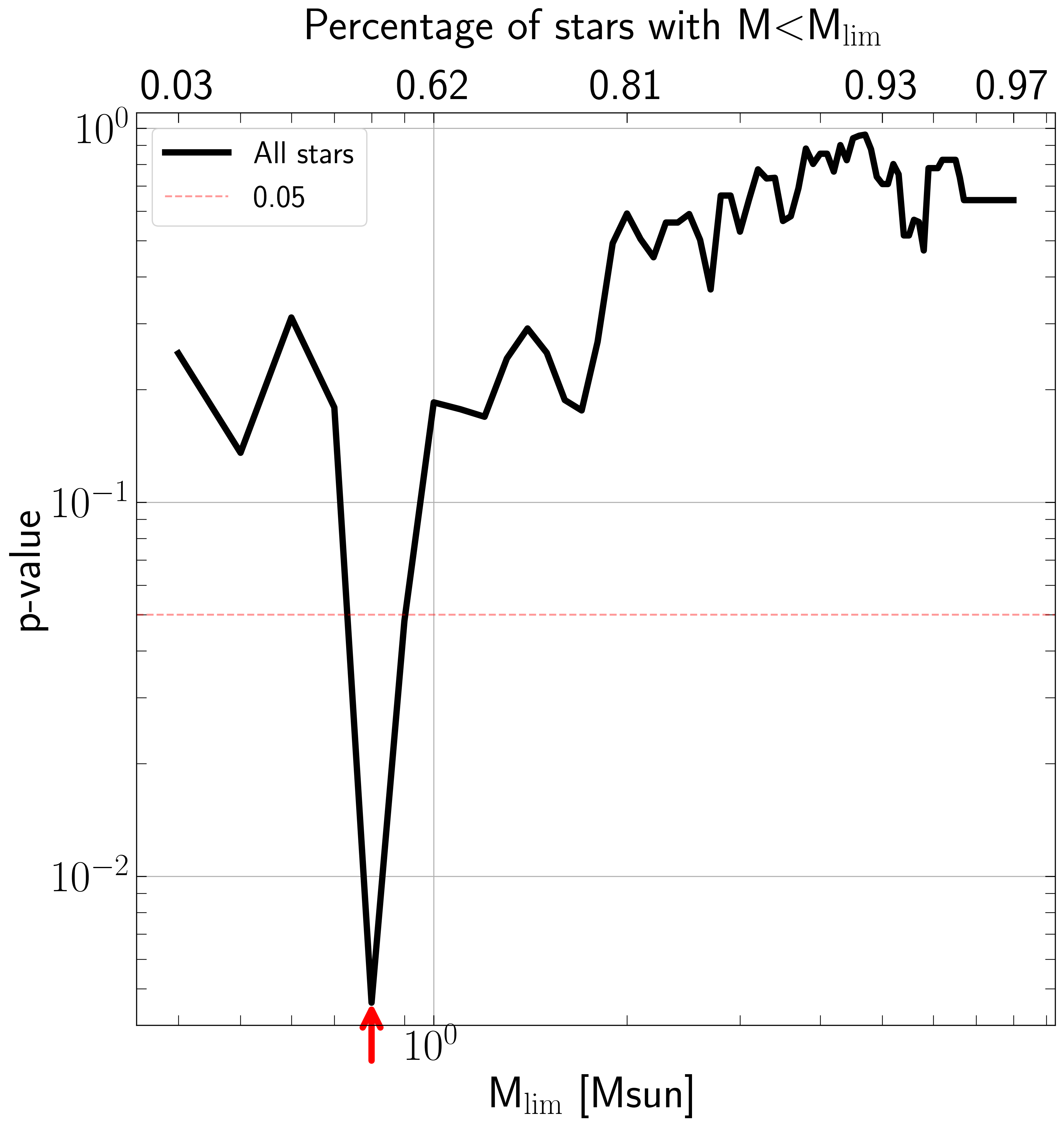} 
\includegraphics[width=0.66\columnwidth]{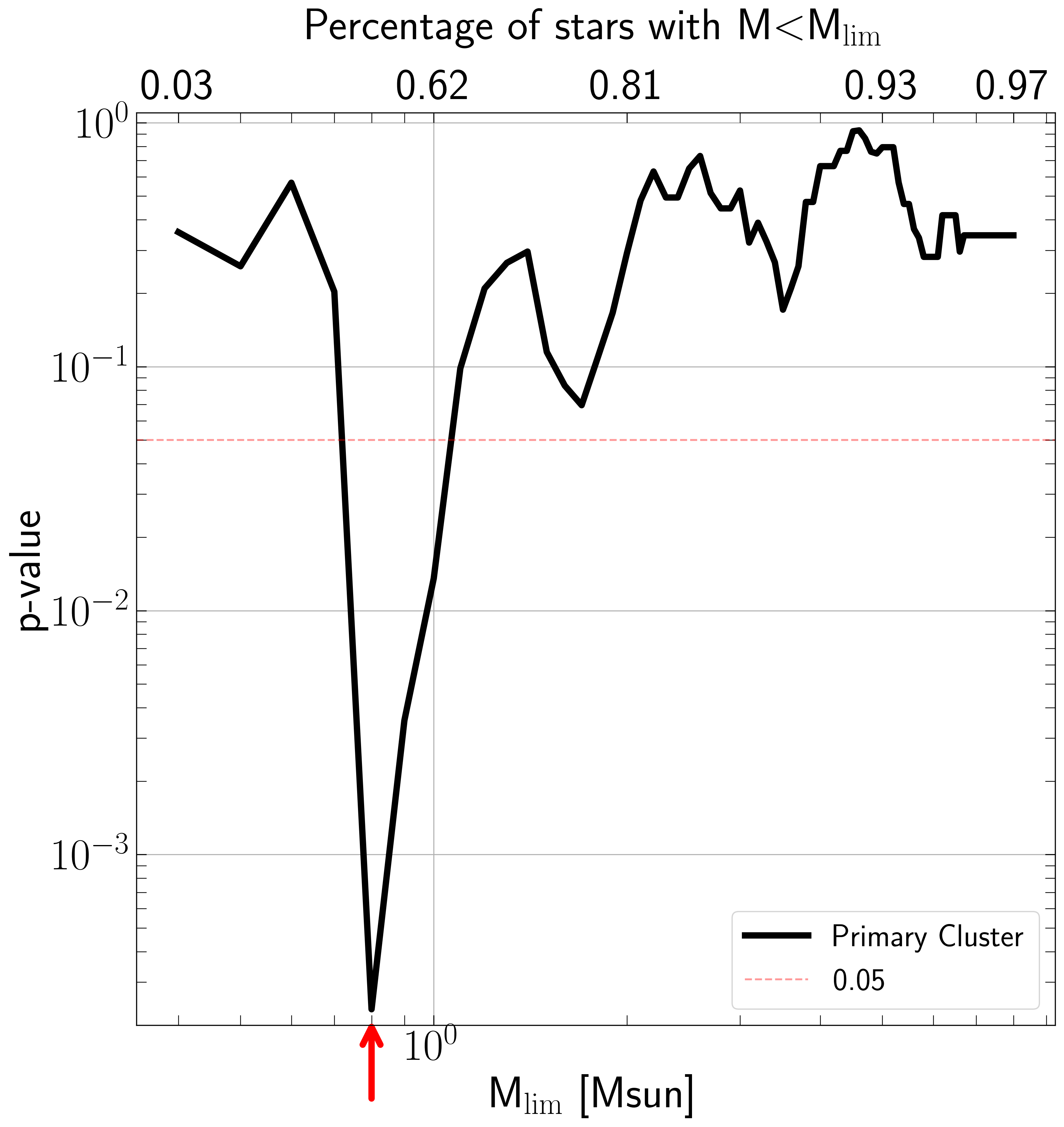} 
\includegraphics[width=0.66\columnwidth]{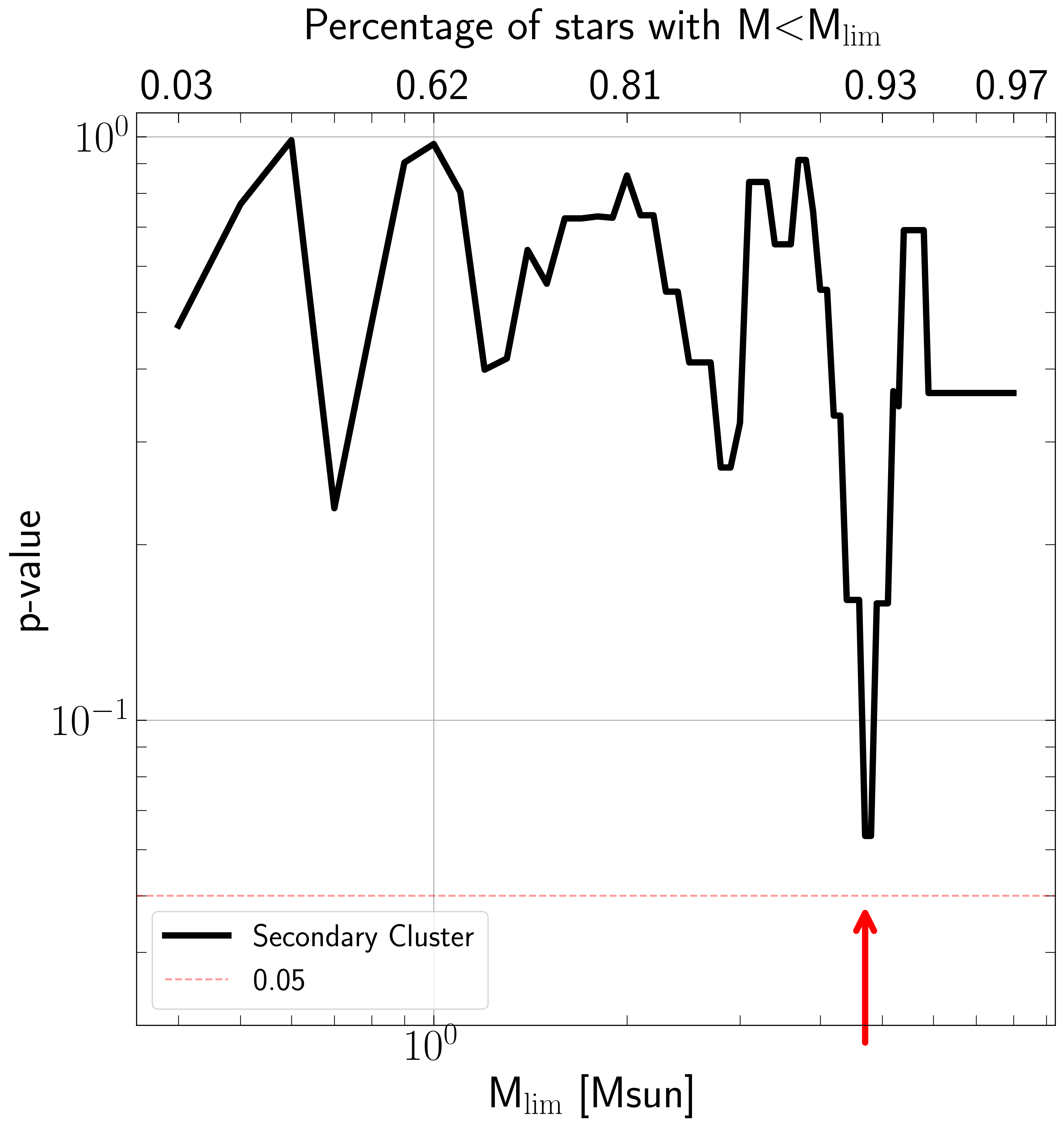}\\
 \caption{
$p-$values of the KS-test as a function of $\mlim$, between the spatial distribution of massive ($M>\mlim$) and low mass ($M<\mlim$) stars, for stars in Sample~1B. Panel (a), all stars in the whole sample. Panel (b), stars in the Primary group. Panel (c) stars in the Secondary group. The horizontal dotted red line is located at $p=0.05$. The upper $x-$axes indicate the percentage of stars with masses below $\mlim$, for five different values of $\mlim$. The red arrow on the x-axis points to $\mlim$ for which $p$-value $\lesssim$~0.05.}
    \label{fig:KS}
\end{figure*}

The results of this procedure are plotted in Fig.~\ref{fig:KS}, where we show the KS-test's $p-$value ($y-$axis) between the spatial distributions of stars with mass $M>\mlim$ and $M<\mlim$, as a function of $\mlim$ (lower $x-$axis). The upper $x$ axis, furthermore, denotes the percentage of stars with masses below $\mlim$ for a set of 5 values of $\mlim$. 

The horizontal red line indicates $p-$value = 0.05, below which we would reject the hypothesis that the two data sets come from the same intrinsic distribution function. 
{As it can be seen,} with $p>0.05$ for most values of $\mlim$, there is insufficient evidence to suggest that massive and low-mass stars have intrinsically different spatial distributions, which would indicate spatial mass segregation. \\

\begin{figure*}
\includegraphics[width=0.66\columnwidth]{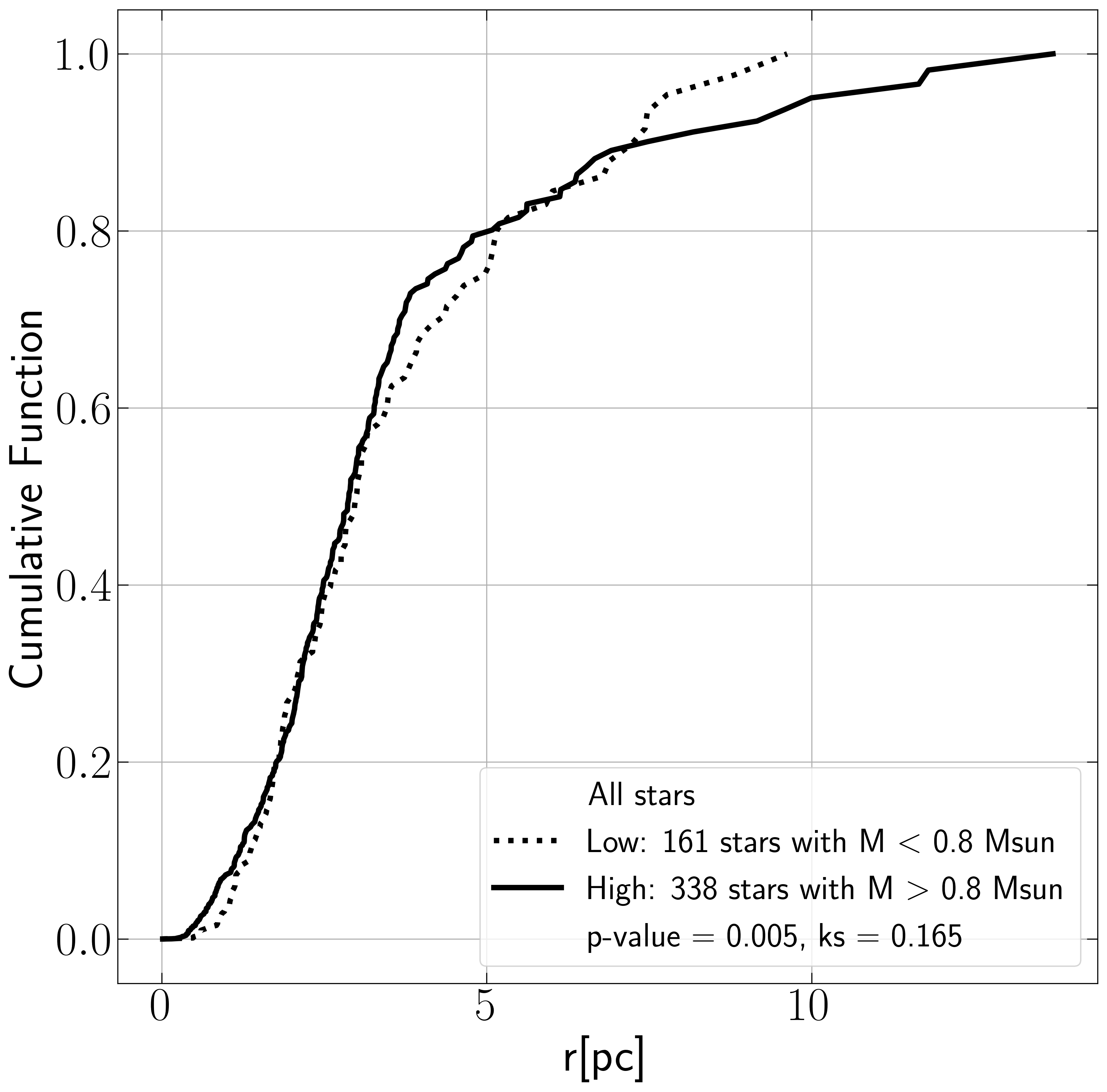}
\includegraphics[width=0.66\columnwidth]{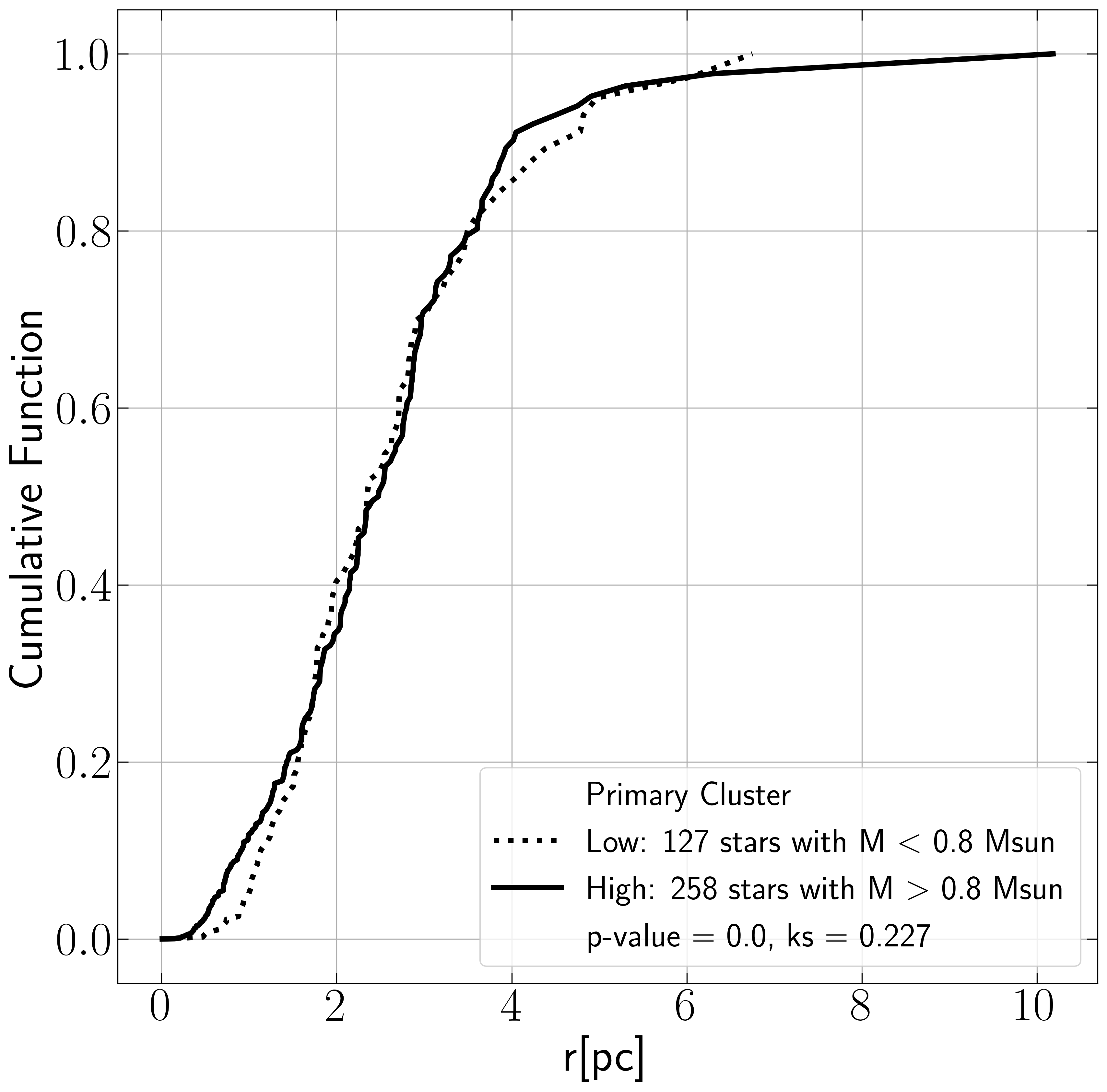}
\includegraphics[width=0.66\columnwidth]{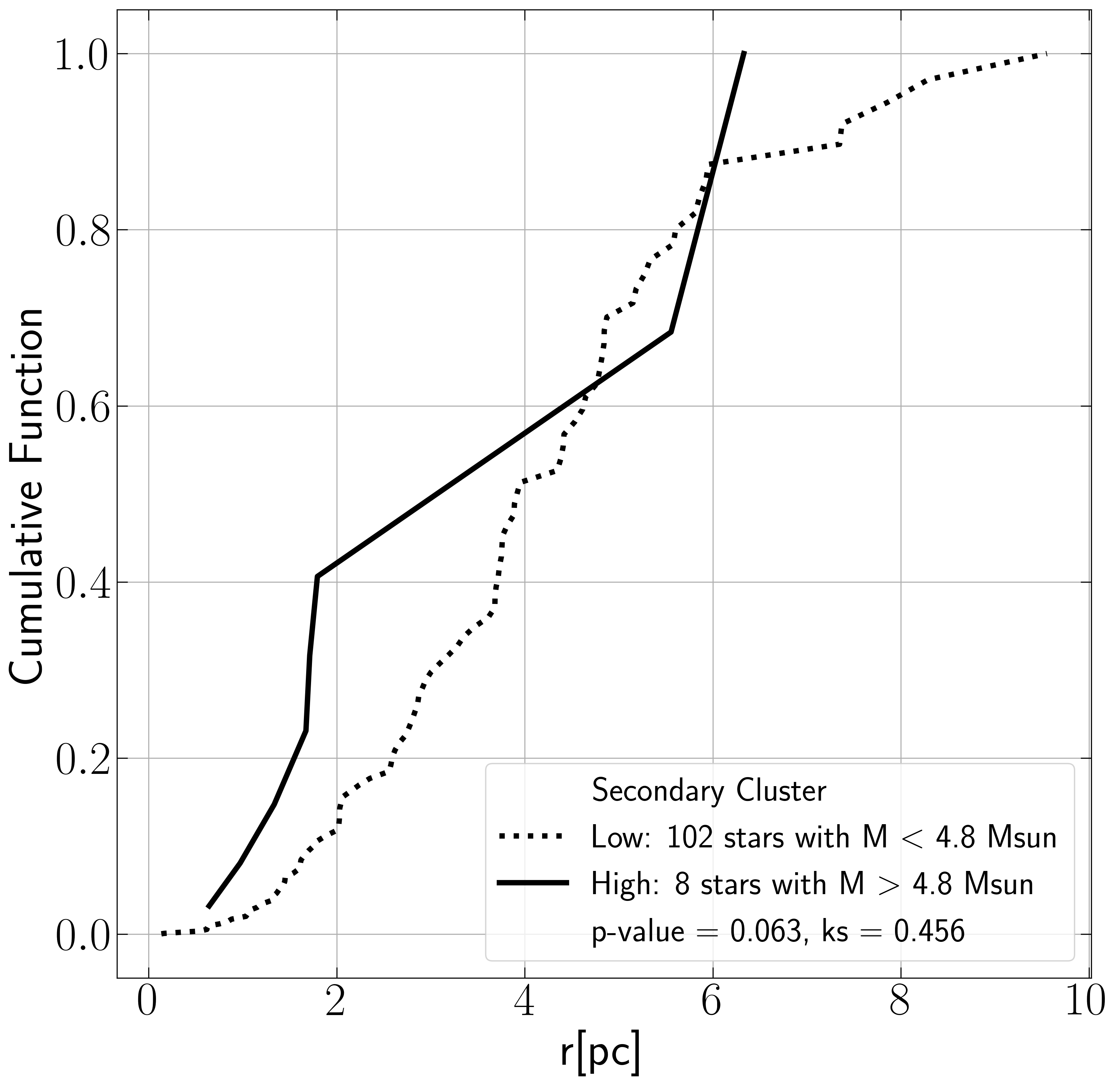}
 \caption{Comparison of the cumulative distributions of high-(solid line) and low-mass (dotted line) stars in the LNC and its substructures. The right, middle, and left panels exhibit the distributions for the whole Sample~1B, Primary, and Secondary Groups, respectively. We used $\mlim=$~0.8, 0.8, and 4.8~\Msun for these distributions, respectively. These values correspond to the cases where we the $p$-values $\lesssim$0.05 (see Fig.~{\ref{fig:KS}}).}
\label{fig:segregation-cumulative}
\end{figure*}

To further argue that the LNC and their substructures have not undergone dynamical relaxation, in Fig.~\ref{fig:segregation-cumulative}, we show the cumulative distributions of massive (solid lines) and low-mass (dotted lines) stars for the whole Sample~1B (left panel), the Primary Group (middle panel), and the Secondary Group (right panel). The distributions shown correspond to $\mlim=$~0.8, 0.8, and 4.8 ~\Msun\ for the left, middle, and right panels, respectively. These are the values for the worst-case scenarios denoted by a red arrow in Fig.\ref{fig:KS}, i.e., by those cases where the $p$-value is minimum ($p\lesssim$~0.05). In Sample~1B and the Primary Group (left and middle panels), the cumulative functions show minimal differences between the distributions of high-mass and low-mass stars, and thus, it is clear that there is no larger concentration of massive stars towards the center of their respective groups, compared to the low-mass stars. As for the Secondary Group (right panel), judging from this figure, it appears to be a larger concentration of massive stars towards the group's center.  \\

\begin{figure*}
\includegraphics[width=0.66\columnwidth]{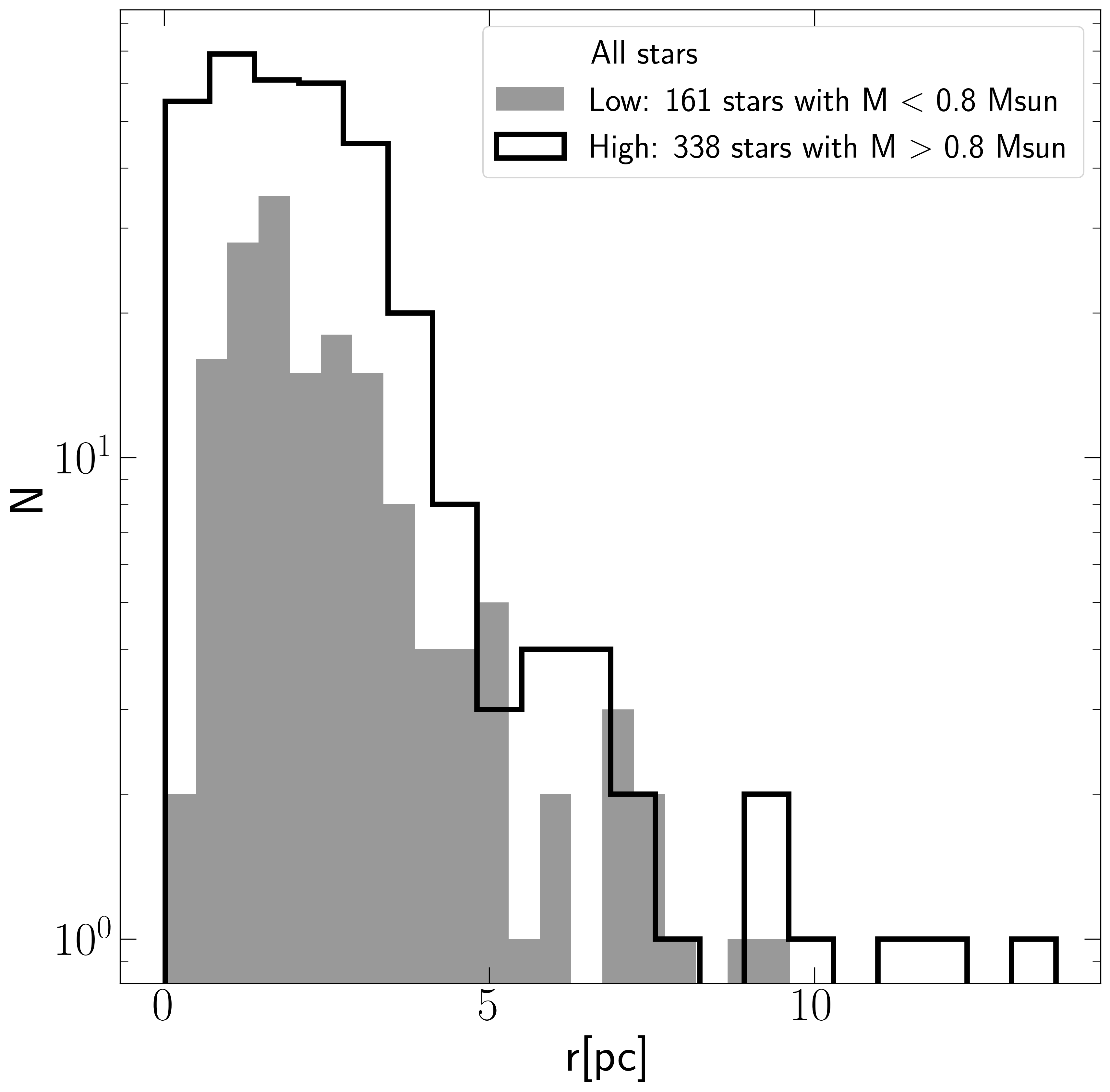}
\includegraphics[width=0.66\columnwidth]{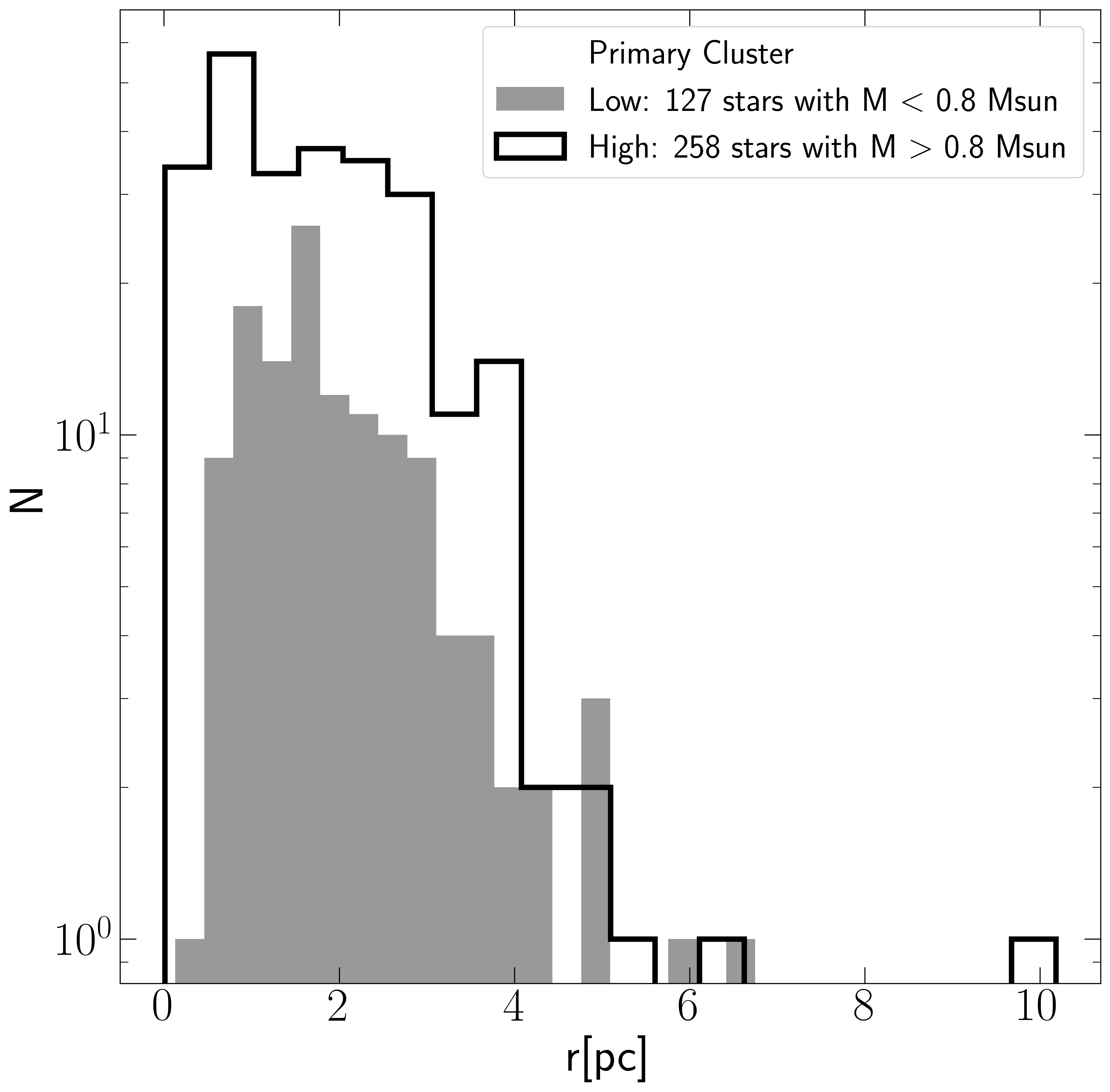}
\includegraphics[width=0.66\columnwidth]{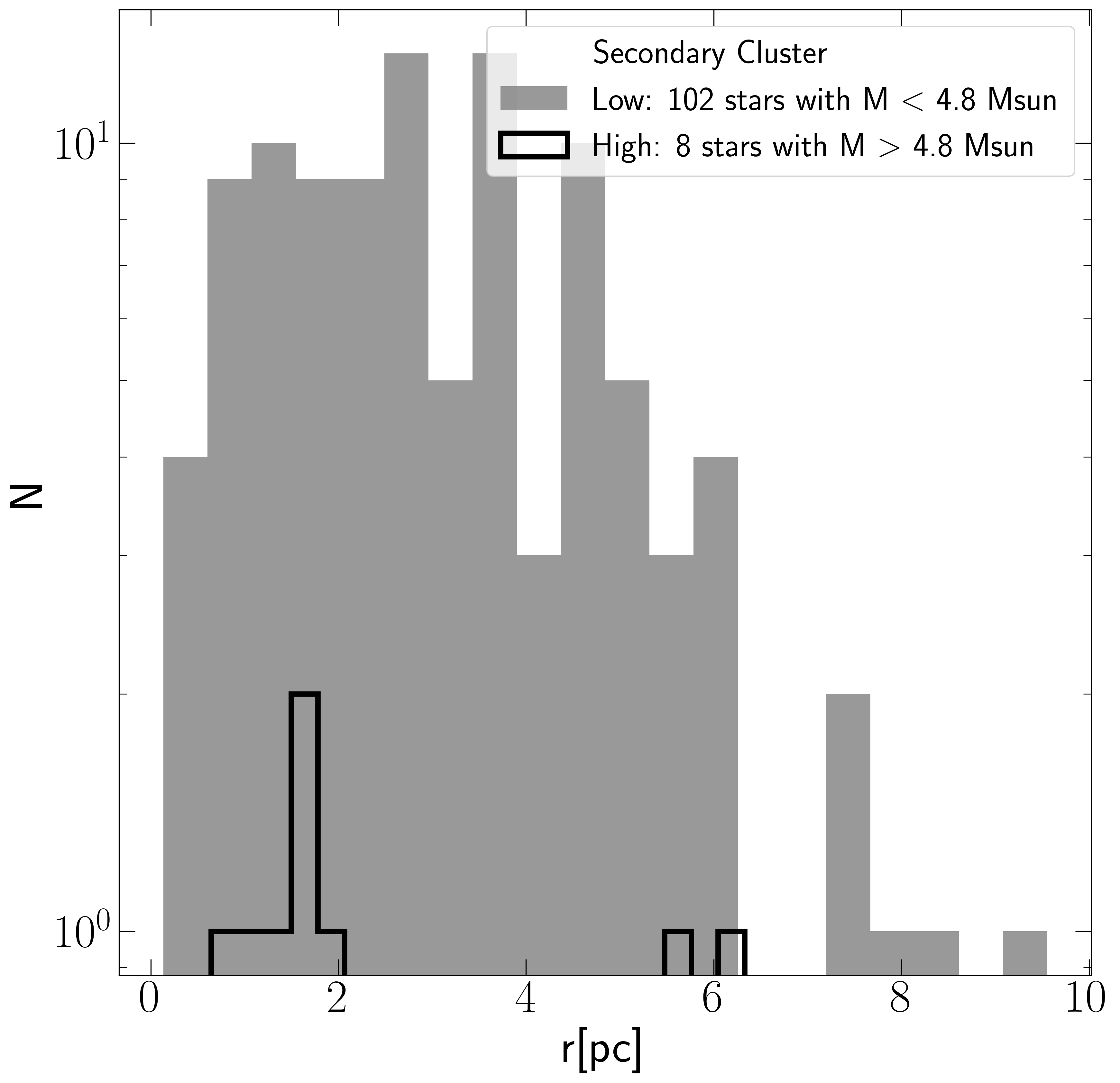}
 \caption{Differential distributions of high- and low-mass stars in the LNC and its substructures for the cases shown in Fig.~\ref{fig:segregation-cumulative}. As it can be seen, it is clear that the low- and high-mass stars in Sample~1B and the Primary Group may very well be drawn from the same intrinsic distribution. As for the Secondary group, even though Fig.~\ref{fig:segregation-cumulative} suggests mass segregation, it is clear that such a result may be spurious due to low-number statistics (only 8 high-mass stars).
}
\label{fig:segregation-histo}
\end{figure*}

To further understand these results, in Fig.~\ref{fig:segregation-histo}, we show the histograms of the high- (open histogram) and low-mass stars (gray histogram) corresponding to the cases shown in Fig.~\ref{fig:segregation-cumulative}. As it can be seen, the whole Sample~1B and the Primary Group show no clear differences between spatial distributions of the high- and low-mass stars, consistent with the discussion above. As for the Secondary Group, we notice that, although Fig.~\ref{fig:segregation-cumulative}c shows signs of segregation, there are only 8 high-mass stars, and thus, one cannot draw conclusions with such poor statistics. \\

As a summary from the previous discussion, we conclude that there are no signs that the whole group, or its subgroups, are mass-segregated for the following reasons: (a) Only in a few cases the $p$-values of the KS test are below 0.05, indicating that, typically, it cannot be said that the low- and high-mass stars are drawn from different distributions. (b) Even in the case of the Secondary Group, which has a $p$-value slightly above~0.05 and its cumulative histogram of the high-mass stars is more centrally concentrated than that of the low-mass stars, the group has too few high-mass stars to draw conclusions.  Thus, we can argue that, in general, there are no signs of dynamical relaxation in the LNC.\\

\subsection{No evidence of massive stars undergoing dynamical heating}

\begin{figure}
 \includegraphics[width=0.9\columnwidth]{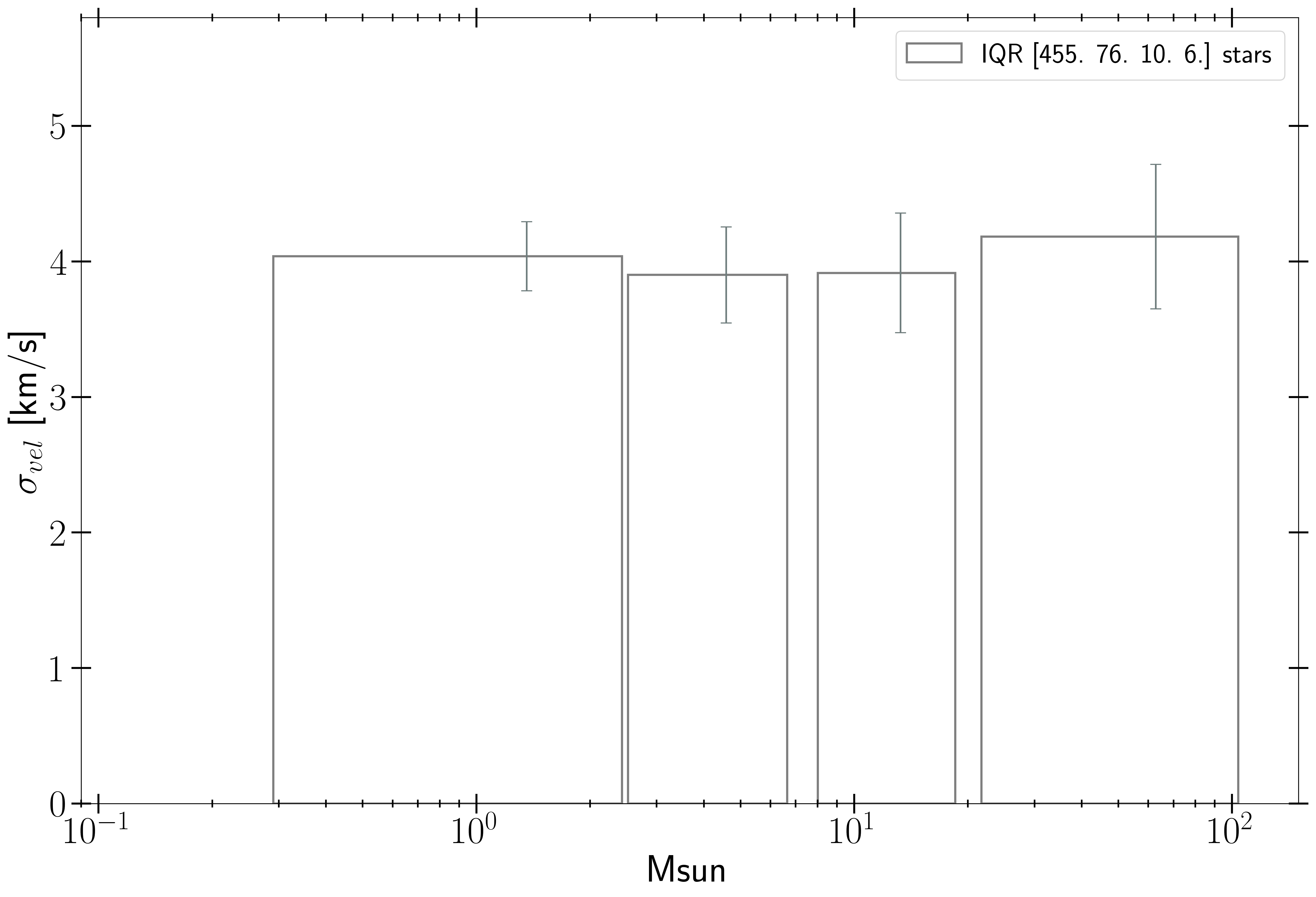}  
 \includegraphics[width=0.9\columnwidth]{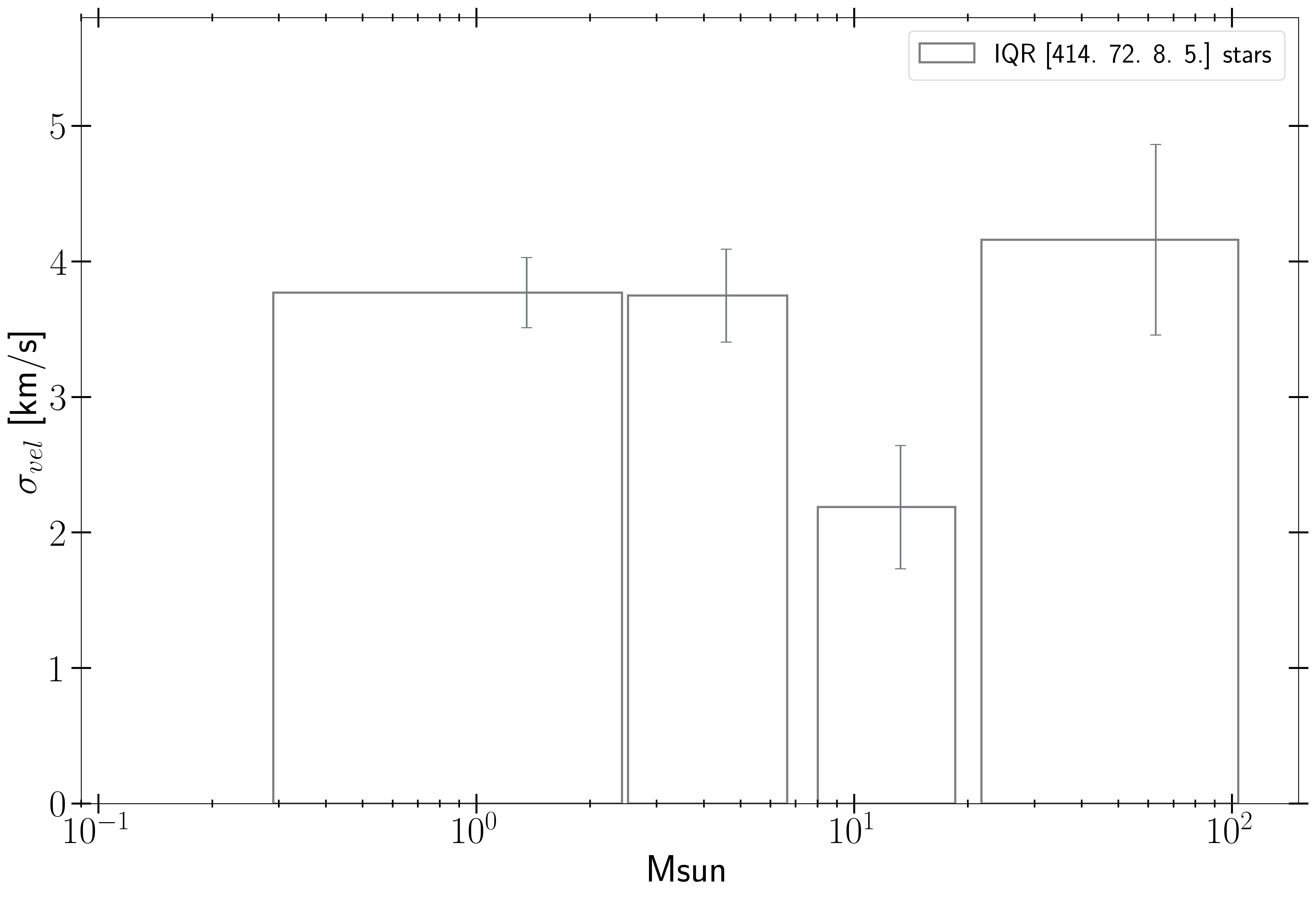}  
 \includegraphics[width=0.9\columnwidth]{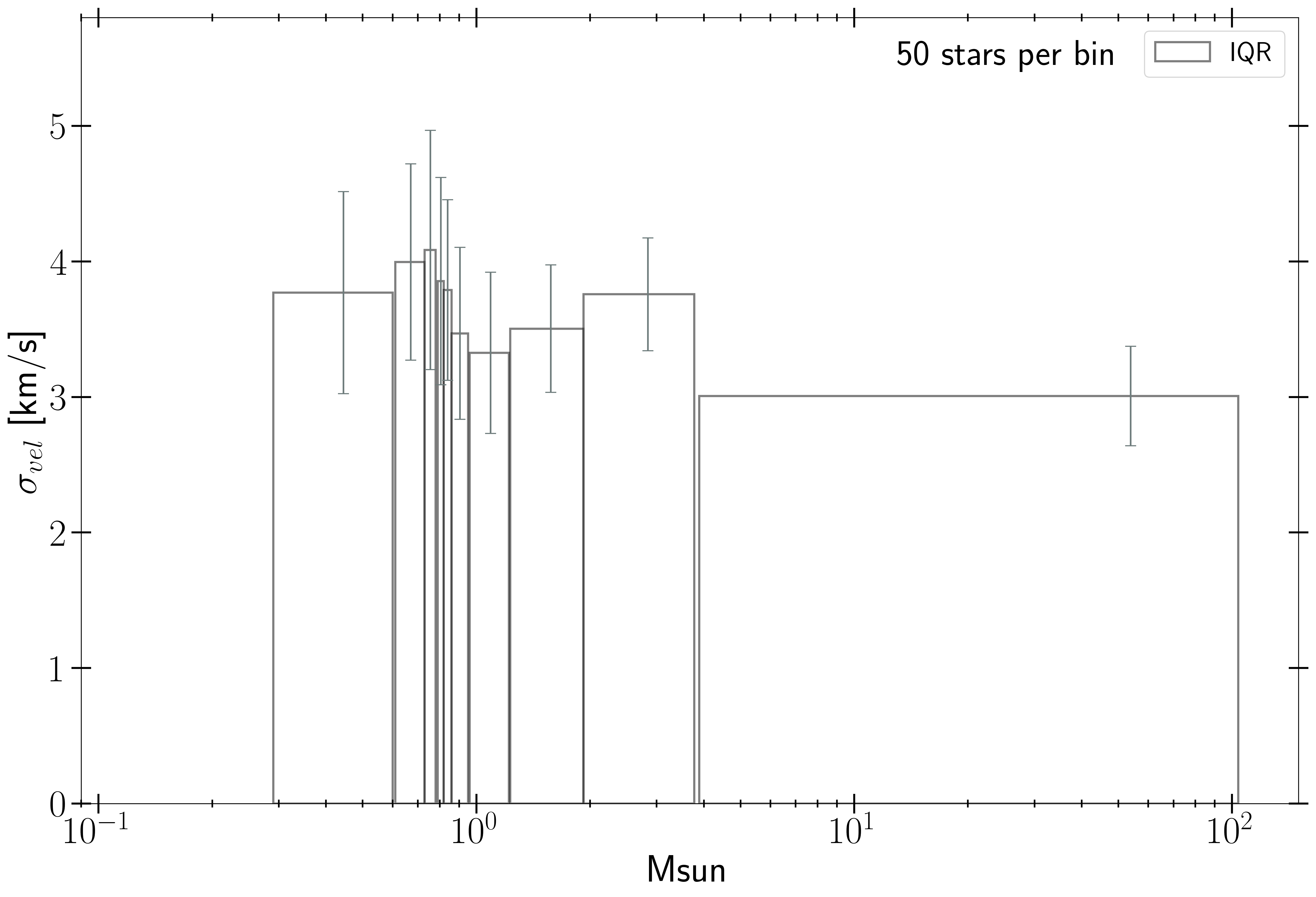}  
  \caption{Velocity dispersion per mass bin for (a) Sample~1A. (b) Sample~1b. Both cases have the same bins as \citet{Wright_Parker19}. Panel (c) shows Sample~1B with equal number of stars per bin. 
  }
    \label{fig:dv-mass}
\end{figure}

Since we want to understand whether the massive stars in the LNC are undergoing dynamical heating {\color{black} as the result of collisional relaxation}, we first want to verify whether the dynamical heating reported previously with Gaia DR2 data remains in Gaia~DR3. Therefore, in Fig.~\ref{fig:dv-mass}~a, b, we show\footnote{It is worth recalling that these plots are not histograms, i.e., these are not number per mass bin, but velocity dispersion-mass plots. We show them as histograms because in order to compute a velocity dispersion, it is necessary to take a mass-bin.}, using the same masses and mass bins as \citet{Wright_Parker19}, the velocity dispersion per mass bin for the stars in (a) Sample~1A and (b) Sample~1B. In contrast, Panel (c) shows the velocity dispersion of Sample~1B with equally spaced mass bins.  \\

We note that the velocity dispersion as a function of the mass bin is substantially flatter than the one reported by \citet{Wright_Parker19} in all three panels. A striking feature at first glance is that the second most massive bin has a substantial drop between panel (a) and (b). However, the reason for this discrepancy is just a poor-statistics effect. Panel (a) has ten stars in that bin, while panel(b) has eight, making unreliable the statistics provided with such low numbers. This is precisely the reason for including panel (c): to avoid low-number statistics by plotting Sample 1~B with the same number of stars in each mass bin. From this panel, it is clear that the velocity dispersion does not increase as a function of the mass bin of the stars. Instead, it looks substantially flat.\\



In order to furthermore understand the large velocity dispersion of the massive stars in the original sample by \citet{Wright_Parker19}, we recall  Fig.~\ref{fig:punto-vector}, where we showed the vector-point diagram, i.e., the proper motion in right ascension {\it vs.} proper motion in declination, for the stars in Sample~1A (green) and 1B (yellow). As commented before, the green points do not pass the tests given by astrometry quality described in \S\ref{sec:sample1} and have velocities substantially different than those of the whole cluster. In addition, we show with star symbols the six most massive stars in Sample~1A, i.e., those stars with masses larger than 20~\Msun, according to \citet{Wright_Parker19}. From this plot, it is clear that one of these massive stars has substantially larger proper motions compared to the characteristic proper motion of the LNC. This star is mostly responsible for the large velocity dispersion of the high-mass bin in \citet{Wright_Parker19}. \\

At first glance, one may be tempted to argue that our result is flawed from origin: our $3\sigma$ cleaning process eliminates precisely the massive star that could have been ejected by {\color{black} collisional relaxation}. However, in order to show that it is unlikely that  {\color{black} the velocity of this star is the result of collisional relaxation} in the LNC, we notice that this star has a proper motion almost perpendicular to its radius towards the cluster's center. This is seen in Fig.~\ref{fig:mapa:w:propermotions}a, where the massive star rejected from Sample~1A to Sample~1B is the green star symbol at {RA$=$271.06~deg, DEC$=-24.18$~deg}. If the large velocity of this star were due to collisional relaxation, it should be located either at the center of the cluster or anywhere else, but with its proper motions radially aligned, as a consequence of its ejection from the center. Since its velocity is almost tangential, it is unlikely that it has been expelled from the center of the cluster due to any sort of dynamical heating (see \S\ref{sec:collisional}).  \\  

Furthermore, we argue that it is unlikely that this star belongs to the cluster. The tangential proper motions of this star correspond to a velocity of $\sim$9~km/sec at the median distance of the LNC. It is located at a distance of {$\sim$6~pc} from the cluster's center. In order to be bound, the mass of the LNC had to be of the order of {2$\times$\diezala5~\Msun}, an unusual mass for an open cluster. This mass is ten times the estimated virial mass of the cluster and 50-200 times larger than the estimations by \citet[][1000~\Msun]{Prisinzano+19}, \citet[][2500~\Msun]{Wright+19} \citet[][4000~\Msun]{Kuhn+15}.\\

From the previous discussion, it seems unlikely that this star belongs to the LNC. An intriguing possibility, however, is whether it has undergone substantial gravitational acceleration from the parental cloud. Indeed, numerical simulations show that as the newborn stars in stellar clusters expel their parental gas, the gravitational potential decreases, accelerating the stars in different directions \citep{Geen+18, Zamora-Aviles+19}. This process is called {\it gravitational feedback} \citep{Zamora-Aviles+19}. A further numerical and observational investigation is necessary, however, in order to estimate whether this effect has a relevant impact on the dynamics of stars in star-forming regions.\\

\begin{figure}
\includegraphics[width=\columnwidth]{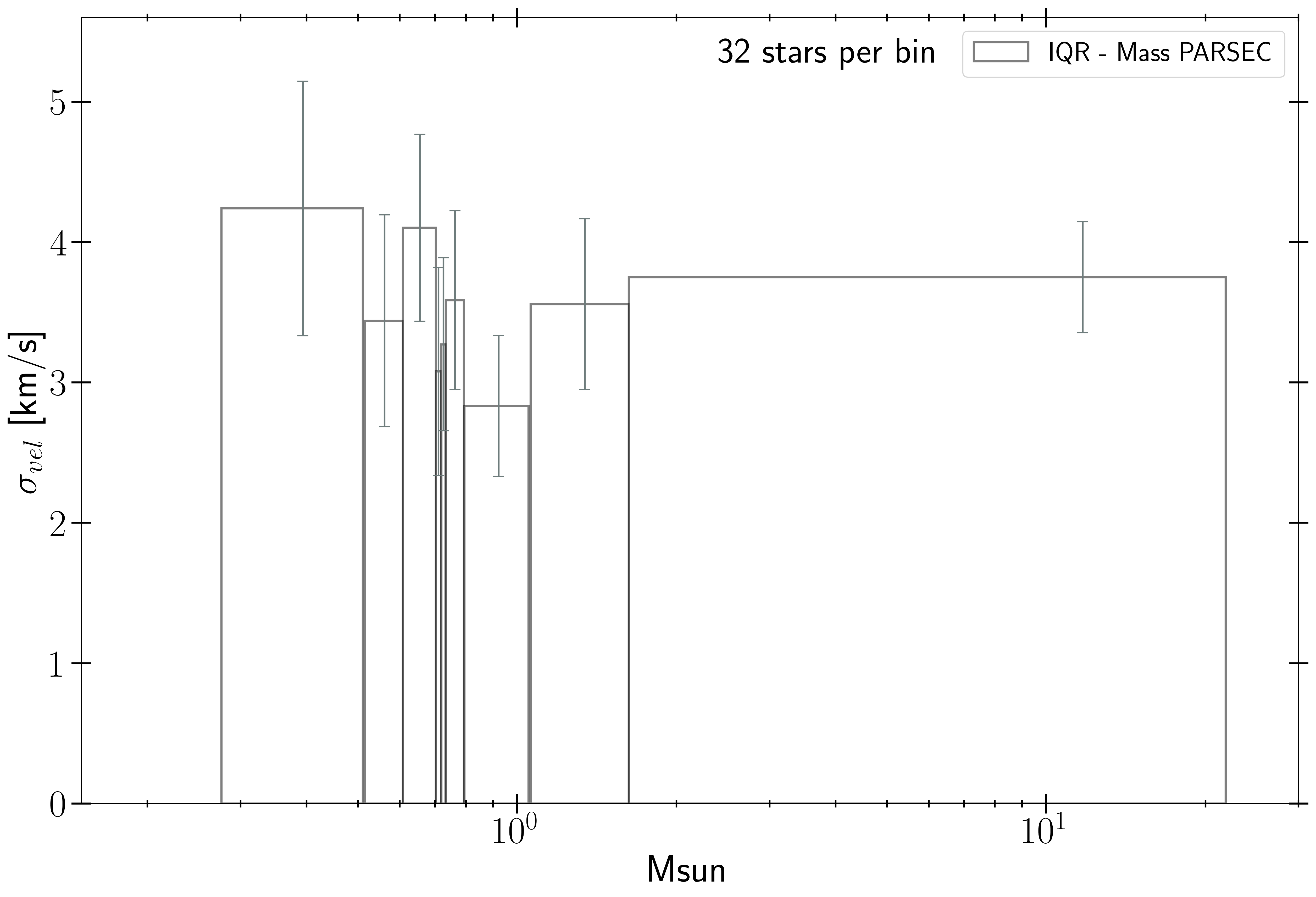} 
\includegraphics[width=\columnwidth]{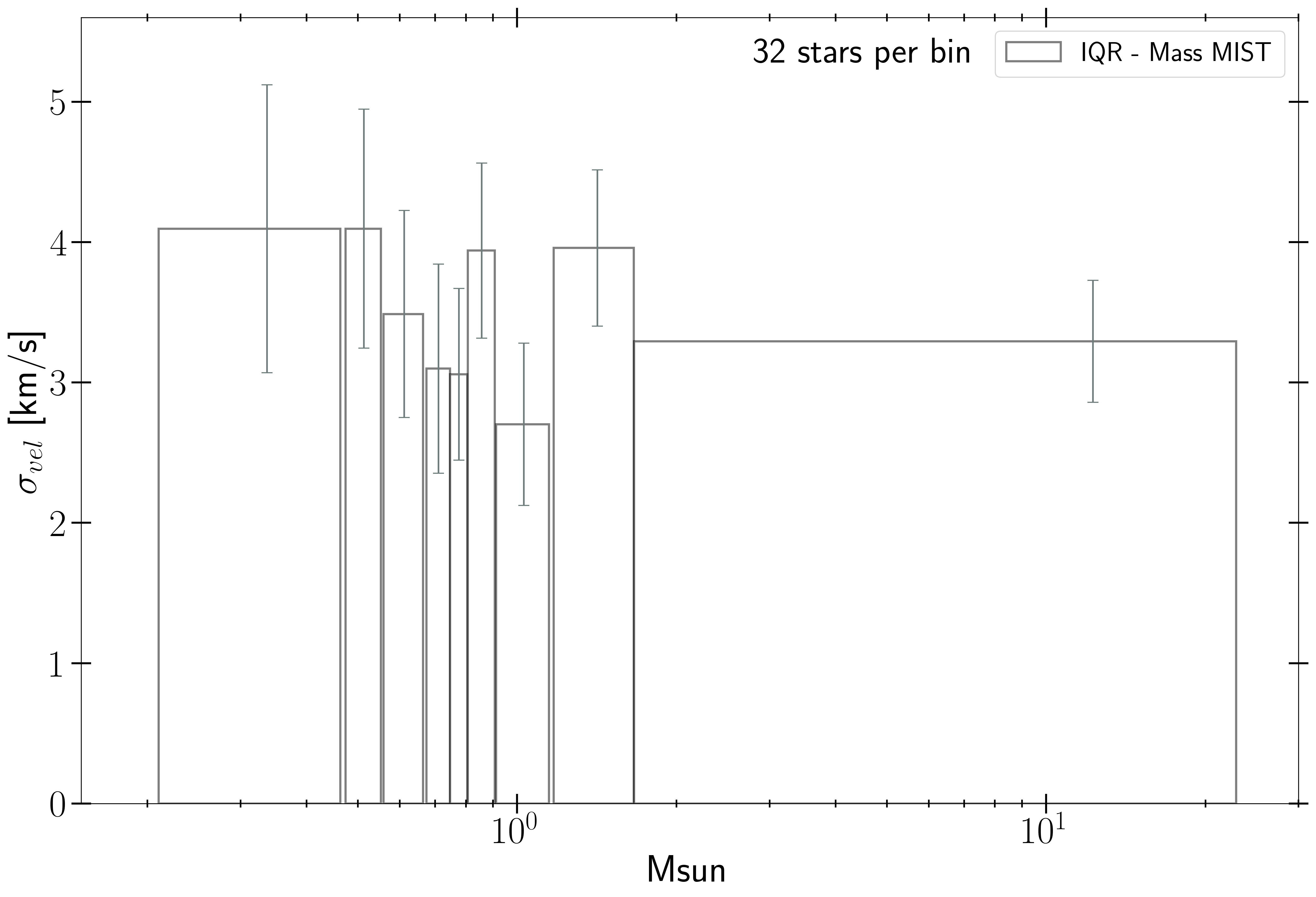} 
 \caption{{Velocity dispersion per mass bin for Sample~2. Masses were computed with \massageparsec. A similar plot is obtained with \massagemist, with no evidence of massive stars undergoing dynamical heating. }}
    \label{fig:sigma-mass:sample2}
\end{figure}

The analysis derived from Fig.~\ref{fig:dv-mass} was obtained using the masses reported by \citet{Wright_Parker19}. 
{It is still necessary, however, to verify whether the masses computed with \massage{} show evidence of dynamical heating.} 
For this purpose, in Fig.~\ref{fig:sigma-mass:sample2}, we first plot the velocity dispersion as a function of the mass bin for Sample~2. Panel (a) shows masses computed with\massageparsec{} while panel (b) shows masses computed with \massagemist. Since, in this calculation, we do not have the same mass ranges as \citet{Wright_Parker19}, we opt to show bins with the same number of stars. This allows us to have similar statistics in all bins. As can be seen, there is no evidence of the dynamical heating of the massive stars.\\

{\color{black}
Similarly, we computed the velocity dispersion per mass bin for the two smaller clusters. The corresponding velocity dispersion-mass plots are shown in Fig.~\ref{fig:sigma-mass:groups}, using the masses from \citet{Wright_Parker19}. Panel (a) corresponds to the Main group, while Panel (b) corresponds to the Secondary group. From this figure, we do not see any sign of dynamical heating in each one of the individual groups.  \\

\begin{figure}
\includegraphics[width=\columnwidth]{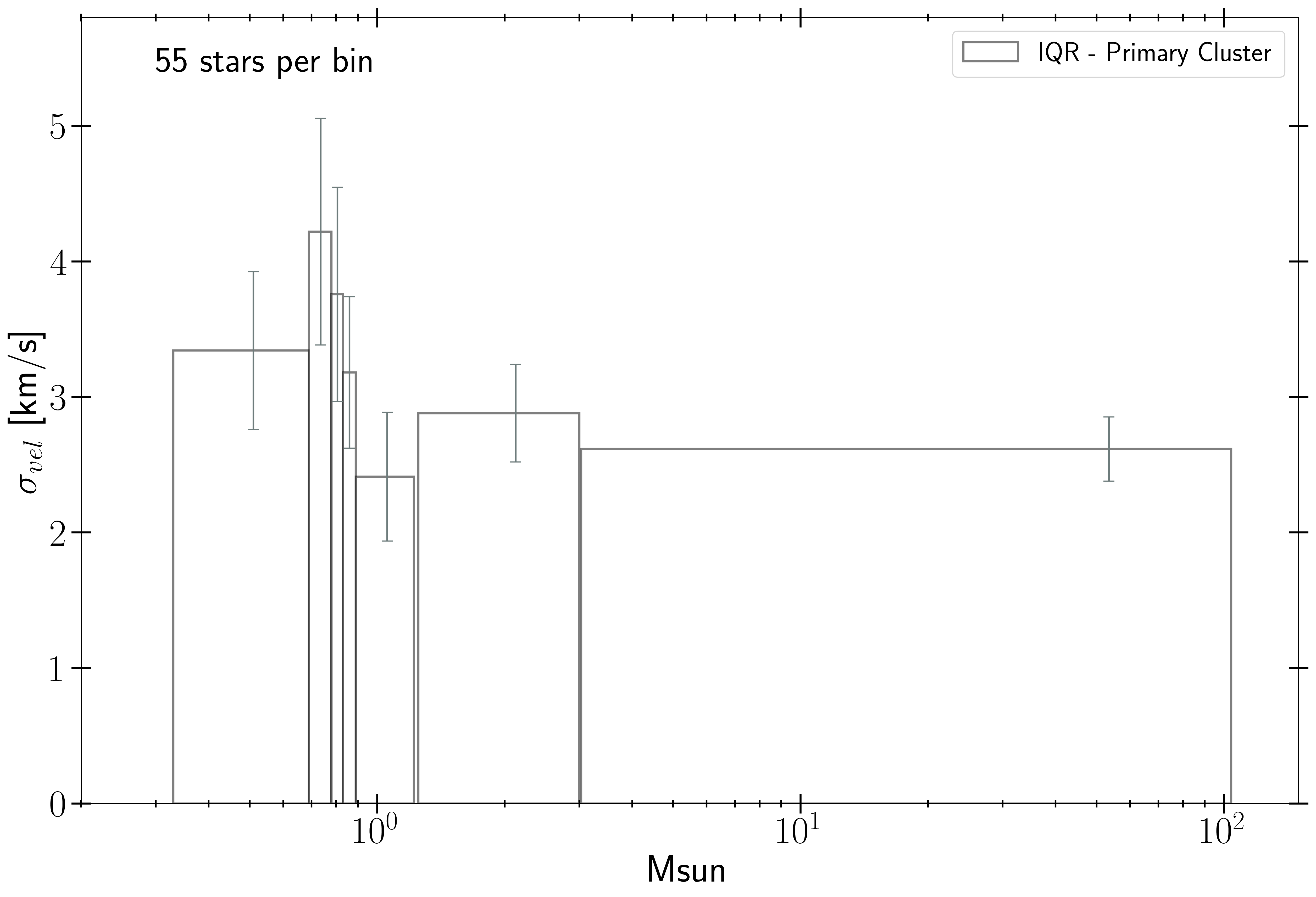} 
\includegraphics[width=\columnwidth]{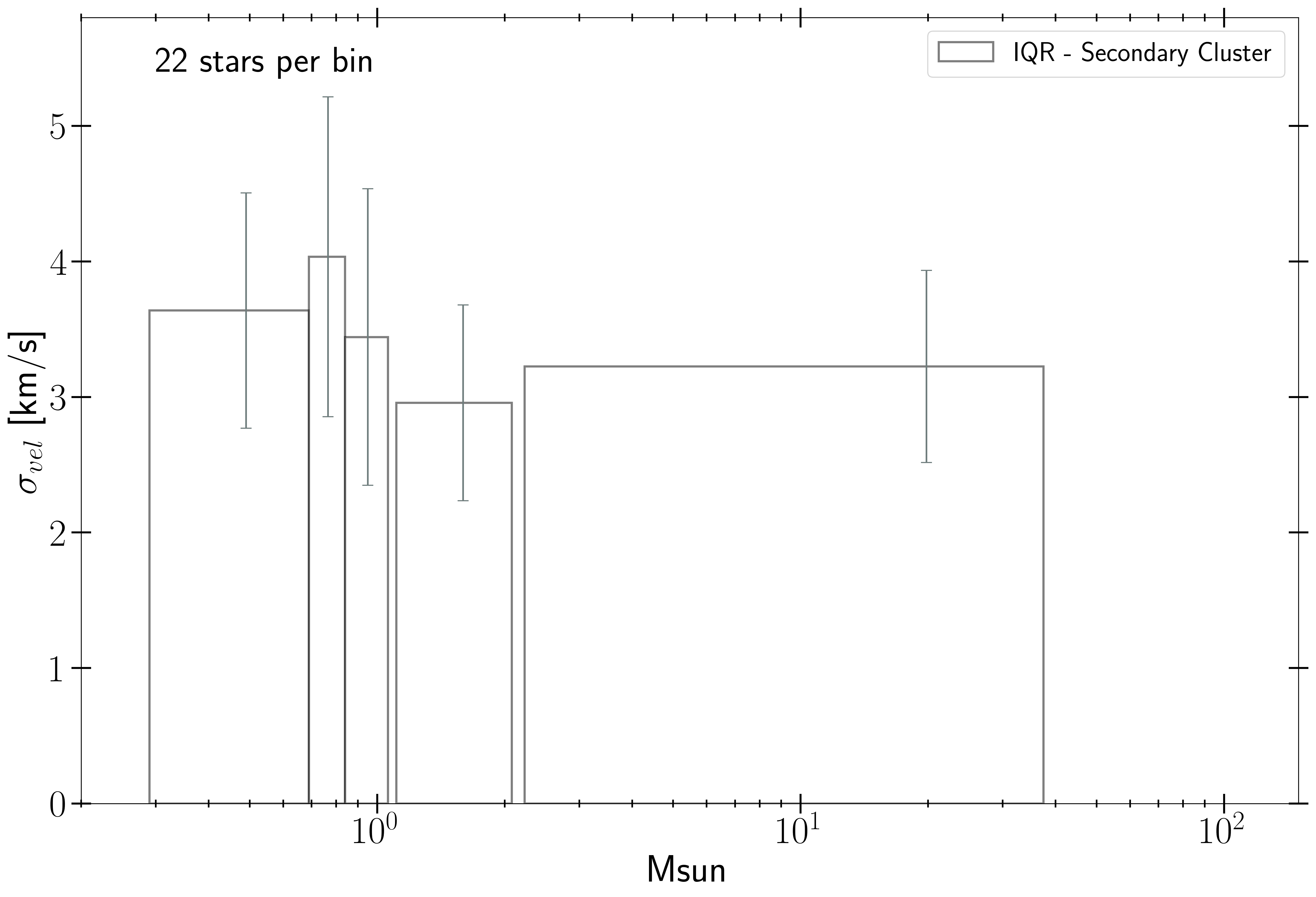} 
 \caption{{Velocity dispersion per mass bin for Sample 1B. Panel (a) Main group. Panel (b) Secondary Group. For these plots we utilize the masses reported by \citet{Wright_Parker19}, in order to have better statistics. }} 
    \label{fig:sigma-mass:groups}
\end{figure}

The situation is different, however, if we use the masses inferred with \massage. Indeed, as can be seen in Fig.~\ref{fig:sigma-mass:groups-massage}, although the Primary group does not exhibit massive stars having larger velocity dispersion than low-mass stars (upper panel), the massive stars in the Secondary group (lower panel) do it. \\

The cause for massive stars' larger velocity dispersion than low-mass stars in the Secondary group is unclear. In dynamical relaxation, one would expect massive stars to not only have larger velocity dispersion but also to be preferentially concentrated towards the center of the group. However, in \S\ref{sec:SpatialSegregation} we found that massive stars are not preferentially concentrated towards the group's center. Thus, it is hard to argue in favor of dynamical relaxation being the cause of the larger velocity dispersion of massive stars in the Secondary group.}

\begin{figure}
\includegraphics[width=\columnwidth]{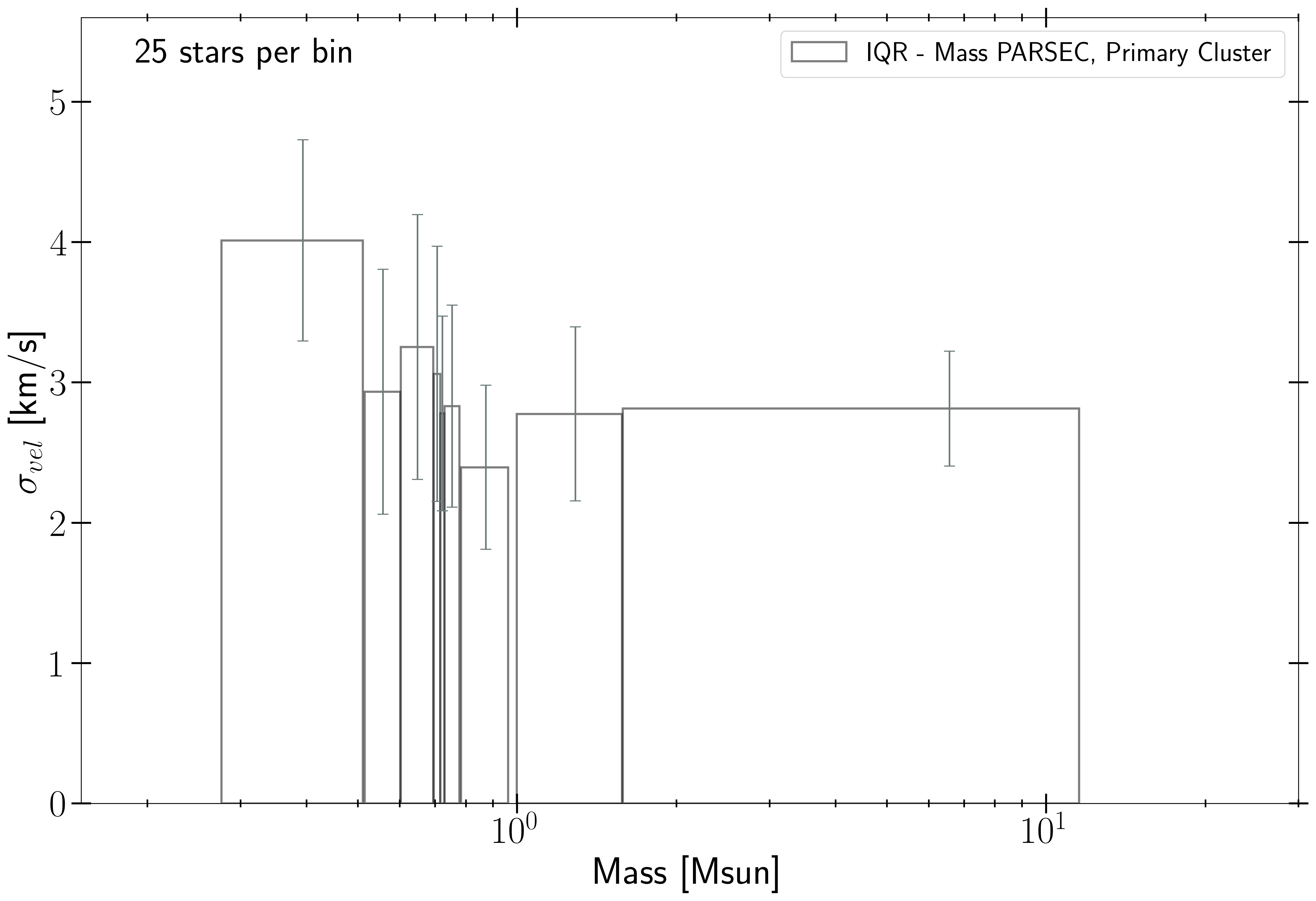} 
\includegraphics[width=\columnwidth]{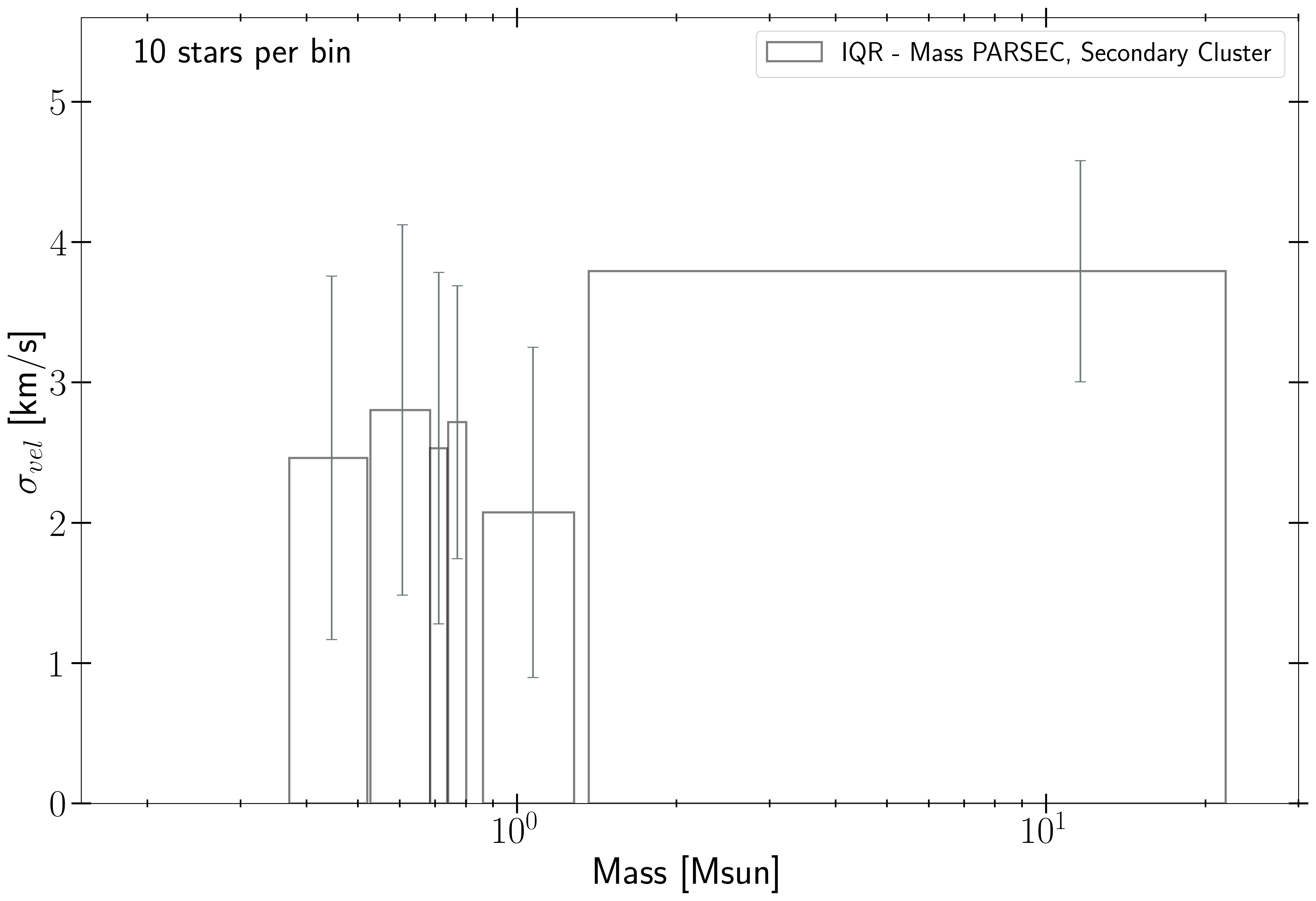} 
 \caption{Velocity dispersion per mass bin for Sample 2 Panel (a) Main group. Panel (b) Secondary Group.} 
    \label{fig:sigma-mass:groups-massage}
\end{figure}

\section{Discussion}\label{sec:discussion}

Accurate astrometrical measurements are crucial to determining the origin and dynamical state of young stellar systems. We started this project aimed at understanding the differences in the kinematic state of the Orion and the Lagoon nebula clusters. In \citet{BonillaBarroso+22}, we found that the former exhibits constant velocity dispersion per mass bin. For the latter,  in contrast, \citet{Wright_Parker19} found that massive stars have larger velocity dispersion than low-mass stars. This discrepancy in their apparent kinematic state can be explained in terms of different dynamical mechanisms dominating the evolution of these clusters. Indeed, although different dynamical mechanisms may be simultaneously at play \citep[e.g., ][]{Krause+20}, this discrepancy suggested that the main dynamical relaxation mechanisms working in the ONC and the LNC should be, primarily, violent and collisional relaxation, respectively. \\

As seen in \S\ref{sec:violentANDcollisional}, the varying character of the gravitational potential during collapse produces violent relaxation. Its main characteristic is that the particles in the system end up with the same velocity dispersion, regardless of their mass. The timescale for this to occur is of the order of the free-fall time. This means that a young cluster formed from a \diezala4 cm\alamenos3  molecular cloud clump will exhibit signs of violent relaxation within 1--2~Myr. In contrast, although collisional relaxation should, in principle, produce energy equipartition between particles,  the final effect is the opposite. Due to the negative heat capacity of gravitational systems, collisional relaxation increases the velocity dispersion of massive stars compared to that of low-mass stars. An extreme case of the collisional relaxation process is the \citet{Spitzer69} instability, where the massive stars become so concentrated that they are detached from the total gravitational potential and evolve at a faster rate. The time evolution for this process is of the order of $\sim$8--10~Myr \citep{Parker+16}, substantially larger than violent relaxation. \\

In addition to the distinctive velocity dispersion of each mechanism, spatial mass segregation may also give clues on the dynamical processes in the clusters. For instance, collisional relaxation necessarily produces spatial mass segregation. In contrast, violent relaxation can occur without it, although it is frequently associated to \citep[see, e.g., ][]{Bonnell+07, Kuznetsova+15}.  Thus, the lack of spatial mass segregation discards collisional relaxation as a mechanism playing an essential role in the dynamical state of a cluster.\\

In the present work, we used Gaia DR3 measurements for the LNC, with the idea of having an up-to-date estimation of its dynamical state. To avoid spurious results, we furthermore rejected stars with significant uncertainties or bad astrometric solutions. As a result, we found that the massive stars in the LNC do not have larger velocity dispersion compared to low-mass stars. In addition, we also did not find evidence of spatial mass segregation.  Thus, we conclude that the LNC shows no signs of collisional relaxation. \\

In our quest to infer its dynamic state, we found that the LNC is composed of two expanding stellar groups overlapping. The bigger group has a dynamical age of $\sim$1.4~Myr, while the smaller one $\sim$1.47--3.89~Myr, and is located in the western part of the large group. Their centers of mass are approaching each other at a velocity of $\sim$ 3$~\kms$. \\



We also did not find evident spatial and velocity mass segregation of the Primary group. Instead, we found that {\color{black} stars are well-mixed and exhibit constant velocity dispersion per mass bin. Interestingly, although there is no spatial mass segregation in the Secondary group, massive stars have larger velocity dispersion. At first glance one can be tempted to think that this may be produced by massive binary stars having larger orbital velocities. However, our selection of stars with RUWE values smaller than 1.4 rules out this possibility. We thus speculate} a possible mechanism that may have produced this behavior. On the one hand, this group may have suffered some dynamical relaxation, first, and then some sort of violent mechanism dispersed it. If so, the massive stars may have first sunk into the center, acquiring larger velocity dispersion. After the explosive event, all stars disperse, with the massive ones reaching substantially large distances, {\color{black} erasing the spatial mass segregation that could have occurred due to the dynamical relaxation but retaining their large velocity dispersion}.


Finally, the fact that the Primary and Secondary groups are approaching each other suggests that they have formed from different parent clumps. The larger clump forms the larger stellar group. Its stellar feedback removes the remaining gas $\sim$1.4~Myr ago, leaving an overvirial expanding cluster. Each stellar group is expanding at its own rate, and they both are approaching each other while expanding because they retain the bulk motion of their parent clump. At this moment, these possibilities are just speculation, and further analysis and observations are necessary to understand the origin of the Secondary group. \\

The Lagoon Nebula Cluster is a young cluster at a median distance of $\sim$~1.24~kpc from the Sun. The selected star sample has ages smaller than 10~Myr \citep{Wright+19}. While one can argue that in some cases, this could be enough time for the cluster to undergo dynamical heating of massive stars  \citep[e.g.,][]{Parker_Wright16}, it is unclear that this is the case of the LNC. The global velocity dispersion of the LNC is of the order of 3~$\kms$, while its diameter is about 10~pc, resulting in a characteristic dynamical scale of $\sim$3~Myr.  The discrepancy between the ages estimated from the isochrones in the HR diagram ($\sim$10~Myr) and the dynamical ages ($\lesssim$3~Myr) suggests that still estimating the age of a cluster with $\lesssim$20\%\ uncertainty in parallax is highly inaccurate. The LNC is probably younger than 10~Myr, and its Primary group may have been formed basically by the collapse of a large clump.

\section{Conclusions}

We have analyzed Gaia DR3 data of the Lagoon Nebula Cluster to estimate its actual dynamical state and the masses of the stars in the LNC for which effective temperatures have been estimated elsewhere. Our results can be summarized as follows:

\begin{itemize}
  
  \item The estimations of distances and extinctions to the stars in the LNC that have estimations of the effective temperature give values of distances nearly 100~pc closer and extinctions 0.3 larger than typical values assumed previously.
  
  \item We found that the cluster comprises at least two main expanding groups. The larger one can be associated directly with the whole LNC. The smaller one overlaps the larger one in its western part.
  >
  \item We found no evidence of spatial mass segregation in the LNC or in its two smaller groups. 
  
  \item We found no evidence of velocity mass segregation in the LNC or the Primary group. However, the fact that the Secondary group exhibits a bipolar nature and its massive stars exhibit larger velocity dispersions than its low-mass stars suggests interesting possibilities for the origin and nature of this group that will be investigated elsewhere.
  
\end{itemize}

The fact that the LNC and its Primary group exhibit constant velocity dispersion per mass bin suggests that they have undergone violent relaxation, which favors the scenario of collapse forming the stellar groups. Nonetheless, the fact that the massive stars in the Secondary group exhibit larger velocity dispersion suggests that some degree of dynamical relaxation may occur and that it has to be investigated in the future.

\section*{Data Availability}

The data underlying this paper will be shared on reasonable request to the corresponding author.

\section*{Acknowledgments}

%
A.B.-B. acknowledges scholarship from CONACyT (2022) and UNAM-DGAPA-PAPIIT support through grant number {\tt IN-114422} (2023).
J.B.-P.  acknowledges UNAM-DGAPA-PAPIIT support through grant number {\tt IN-114422}, and CONACYT, through grant number {\tt 86372}.
J.H. acknowledges support from CONACyT project No. 86372 and the UNAM-DGAPA-PAPIIT project IA102921.
J.H and L.A. acknowledge support from DGAPA/PAPIIT grant BG101723 

\bibliographystyle{mnras}
\bibliography{ReferenciasAndrea} 

%



\appendix

\section{On the effect of eliminating stars with the large proper motions. }\label{sec:AppendixRelax}

As mentioned in the text, we are interested in understanding whether the LNC has undergone some degree of dynamical relaxation. Thus, it is important to check that those stars that were eliminated because they had particularly large proper motions were not the hypothetical stars that could have suffered dynamical relaxation because eliminating them will skew our results precisely towards not finding dynamical relaxation in the LNC. \\

\begin{figure}
\includegraphics[width=\columnwidth]{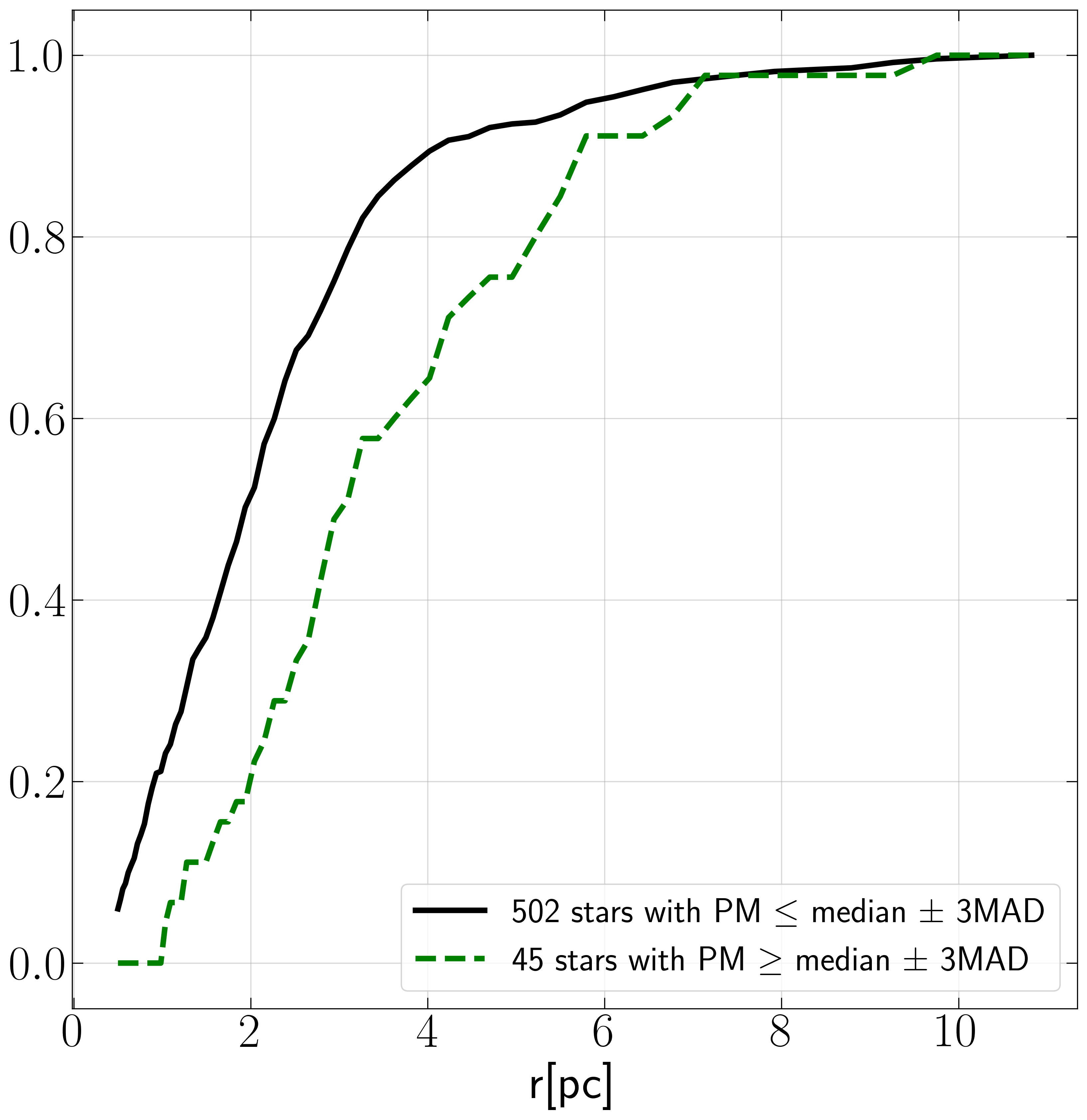}
\includegraphics[width=\columnwidth]{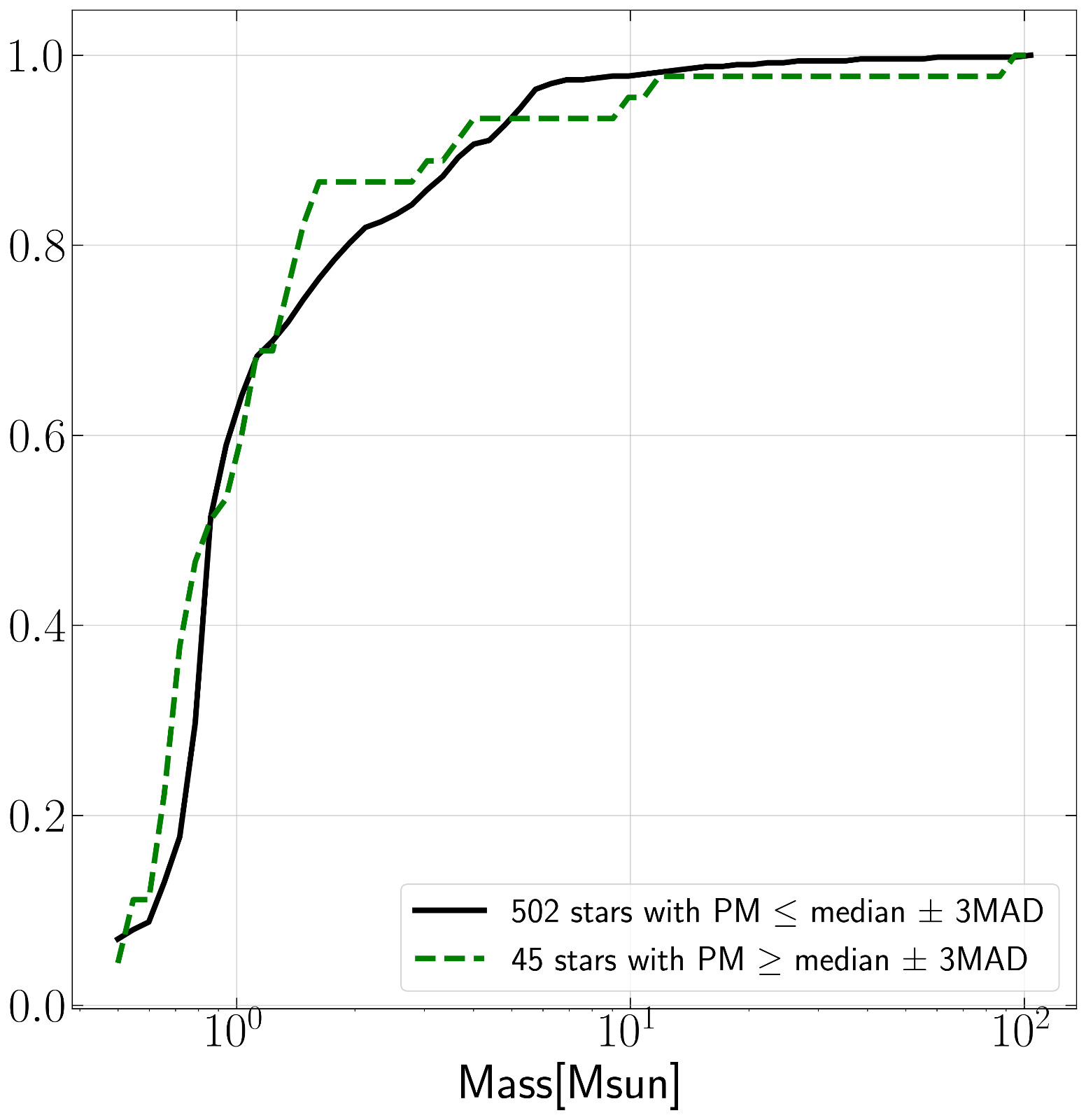}
  \caption{Cumulative distribution of the stars as a function of (a) the distance to the center of the cluster; (b) the mass of the stars. In both panels, green dashed lines correspond to the rejected stars with larger proper motion, and black solid lines correspond to the stars of Sample 1B. As it can be seen, the rejected stars are not particularly more massive, nor are preferentially concentrated in the central part of the cluster, as would be the case if dynamical relaxation had occurred.}
    \label{fig:3madpm:masses}
\end{figure}

We first recall that, due to the collisional relaxation, massive stars move close to the cluster's \textcolor{black}{center of mass}, increasing their velocity dispersion (see \S\ref{sec:collisional}). Thus, we want to investigate whether the eliminated stars are  massive and are preferentially concentrated in the cluster's \textcolor{black}{center of mass}. In Fig.~{\ref{fig:3madpm:masses}}, we plot the cumulative distributions of the 45 rejected stars with large proper motions (green dashed lines) and compare them with the cumulative distributions of the 502 stars in Sample 1~B (black, solid line).  Panel (a) is the cumulative distribution of the positions, while panel (b) is the cumulative distribution of masses of the stars. As can be seen, the 45 stars with large proper motions have no preferentially smaller distances toward the center. On the contrary, they are not at the very center. In addition, these stars are not particularly massive compared to the rest of the population. Thus, we can argue that the rejection step we took in \S\ref{sec:sample1} does not skew our results towards not finding signatures of dynamical heating.

\section{Definition of the smaller groups and their statistical significance}\label{sec:appendix}

At first sight, data points in Fig.~\ref{fig:pos-vel} are grouped into two groups, each one with its own expansion rate and group velocity in the (ra-pmra) plane. In this appendix, we describe how we did assign membership to one, the other, or none of the groups found in  \S\ref{sec:twogroups}. \\

 As can be expected, each cluster-finding algorithm is subject to its methodology, and the outcome will depend on the assumptions and parameters of the algorithm. For instance, if we apply a Gaussian Mixture model to the data, the algorithm will try to find clusters with Gaussian distributions. Similarly, HDBSCAN \citep{McInnes17_HDBSCAN}, one of the most used cluster-finding methods in the literature, will tend to find more roundish clusters. However, the data may not necessarily be Gaussian- or roundish-distributed. This is clearly the present case, where both tendencies in Fig.~\ref{fig:pos-vel} may very well be fitted by a linear correlation. With this idea, we have applied the K-means clustering method in the (ra,pmra) plane.  \\

\begin{figure}
  \includegraphics[width=\columnwidth]{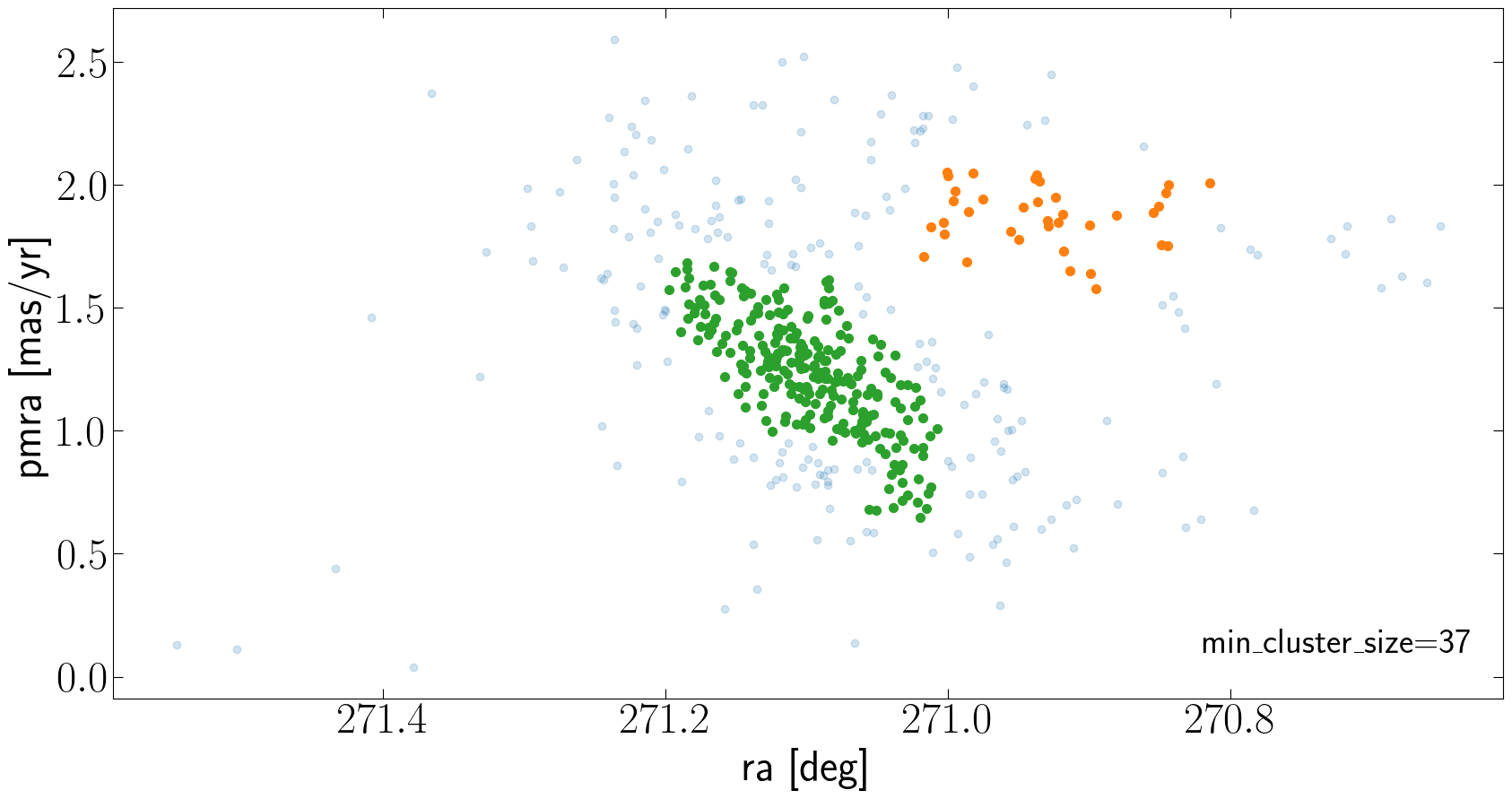}
  \caption{Proper motion in right ascension vs right ascension diagram for stars in Sample 1B. Grey dots represent the whole LNC. Green and orange points are the two groups obtained using HDBSCAN with a minimum of 34 members per cluster. The results are consistent varying the minimum number of members between 21 and 34.} 
  \label{fig:appendix:HDBSCAN:AR-muar} 
\end{figure}

The K-means clustering algorithm \citep{MacQueen67} is an iterative, hierarchical clustering algorithm in which the dataset is split into $n$ groups, and each element of the cluster belongs to the group with the nearest \textcolor{black}{center of mass, or} centroid. Since the initial parameters of the K-means are the number of groups $n$ and the position of their centroids, we have first used the HDBSCAN clustering algorithm \citep[][]{McInnes17_HDBSCAN} to find how many groups there are in the (ra, dec, pmra, pmdec) space, and what their centroids are. The key input parameter of HDBSCAN is the minimum number of members in each group to be found. Thus, we have varied this number between 10 and 40, finding consistent results between 21 and 34. Fig.~\ref{fig:appendix:HDBSCAN:AR-muar} shows the result of using HDBSCAN with a minimum number of 34 members in each cluster. As can be seen from this figure, this procedure left us with two clear groups along the two velocity gradients.  \\  

To determine whether other stars could be assigned to either of the two groups found by HDBSCAN, we applied the K-means method, i.e., we computed the distance from each star to the centroids of each group. We assigned each star to the group whose computed distance is smaller. The distance is computed as the quadratic sum of the normalized right ascension and right ascension proper motion (ra, ~pmra). Once we have assigned membership to one or the other group, we compute the centroid of the each group and repeat the calculation. After three iterations, the stars remained consistently classified within their groups. The resulting groups are named "Primary" and "Secondary", and their (ra,~pmra), (dec,~pmdec) plots are shown in Figs.~\ref{fig:pos-vel:new} and \ref{fig:mapa:w:propermotions}. 

We have left the four stars at the southeast corner of the map unassigned since they do not follow the same velocity gradient and are located more than three times the mean absolute distance from the centroid of the main group. \\

\begin{figure*}
  \includegraphics[width=\columnwidth]{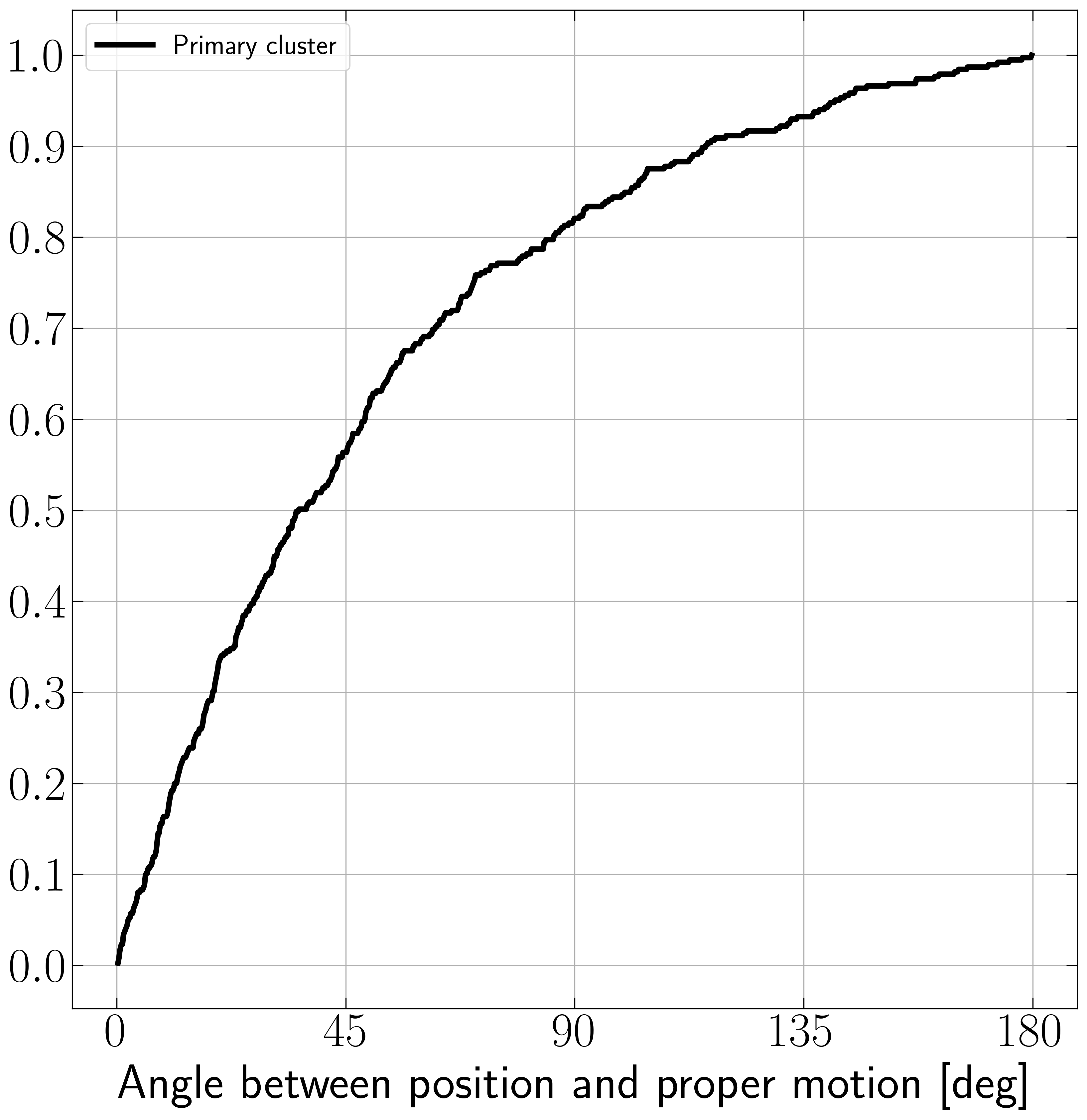}
  \includegraphics[width=\columnwidth]{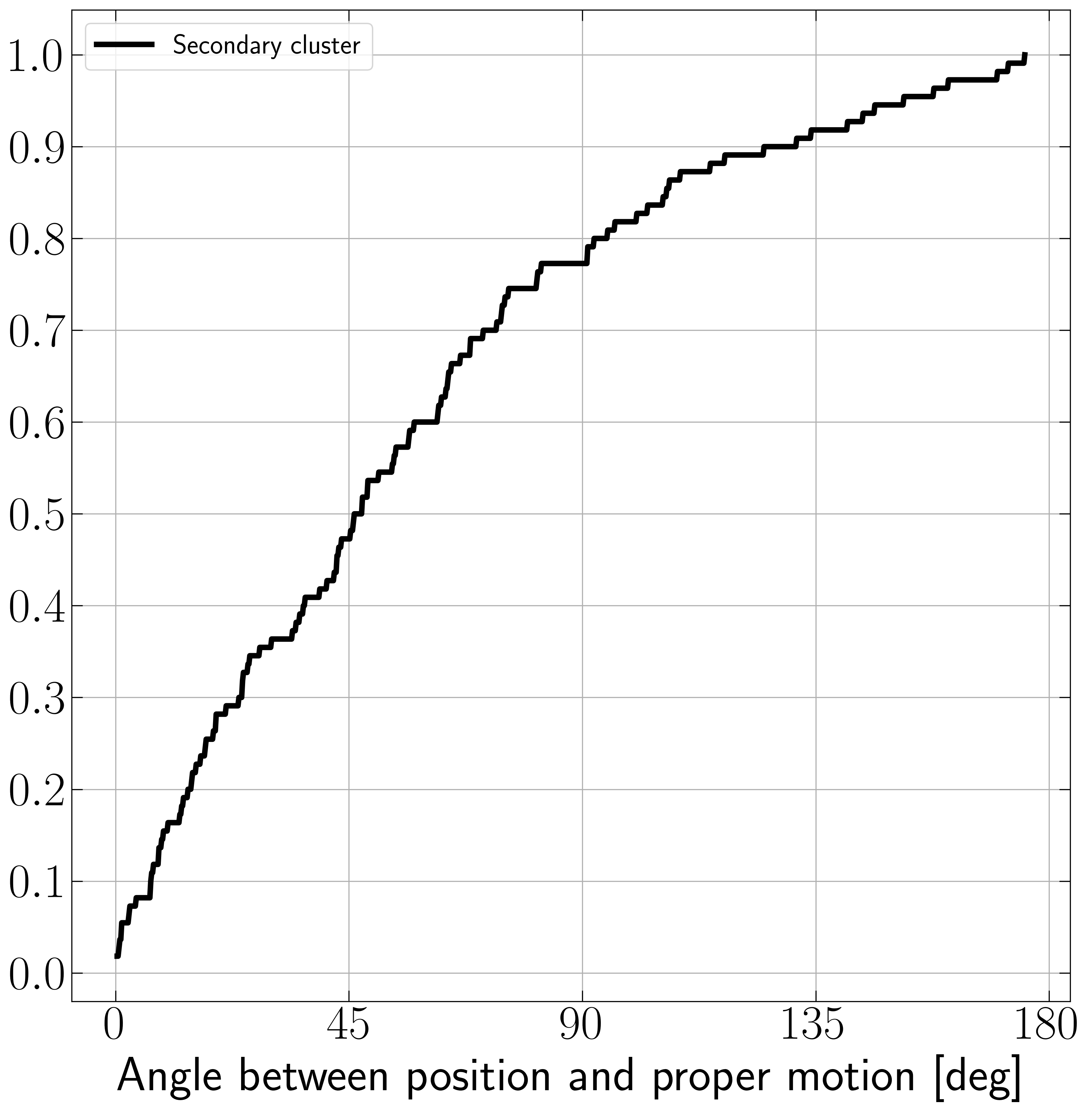}
  \caption{Cumulative function of the angular distribution between the position vector and the proper motion vector, both relative to the cluster center. The left panel represents the Primary cluster, while the right panel represents the Secondary cluster.} 
  \label{fig:appendix:histo:angles} 
\end{figure*}

Before finishing this appendix, we show in  Fig.~\ref{fig:appendix:histo:angles}, the cumulative distributions of the angles between the position vector of each star with respect to the position of the centroid of its group and the proper motion vectors in the frame of reference moving with the mean proper motion of the group. The left and right panels correspond to the Primary and Secondary groups, respectively. As can be seen, $\geq $2/3 of the population of each group has angles smaller than 45$^\circ$, and $\sim$80\%\ has angles smaller than 90$^\circ$, confirming the expanding nature of each group.

\bsp	
\label{lastpage}
\end{document}

%% file: definitions.tex
\def\apj{ApJ}
\def\apjl{ApJL}
\def\aap{A\& A}

\def\mnras{MNRAS}

\def\apjs{ApJS}

\def\mnras{{MNRAS}}

\def\ala#1{$^{#1}$}
\def\alamenos#1{$^{-#1}$}

\def\diezala#1{$10^{#1}$}

\newcommand{\kms}{\rm {km~s^{-1}}}

\newcommand{\mlim}{m_\mathrm{lim}}    %

\newcommand{\Msun}{$M_\odot$}    %
\newcommand{\tauff}{\tau_{\rm ff}}